\documentclass[%
 reprint,
superscriptaddress,
showpacs,preprintnumbers,
 amsmath,amssymb,
 aps,
pre,
]{revtex4-1}

\usepackage{graphicx}
\usepackage{dcolumn}
\usepackage{bm}
\usepackage[normalem]{ulem}
\usepackage{soul,xcolor}
\usepackage{subfig}

\newcommand{\RN}[1]{%
  \textup{\uppercase\expandafter{\romannumeral#1}}}

\begin{document}

\title{Social Clustering in Epidemic Spread on Coevolving Networks}

\author{Hsuan-Wei Lee}
\email{hwwaynelee@gate.sinica.edu.tw}
\affiliation{Institute of Sociology, Academia Sinica, Taipei 115, Taiwan}

\author{Nishant Malik}
\email{nxmsma@rit.edu}
\affiliation{School of Mathematical Sciences, Rochester Institute of Technology, Rochester, NY 14623, USA}

\author{Feng Shi}
\affiliation{Odum Institute for Research in Social Science,  The University of North Carolina, Chapel Hill, NC 27599, USA}

\author{Peter J. Mucha}
\affiliation{Department of Mathematics, The University of North Carolina, Chapel Hill, NC 27599, USA}

\date{\today}

\begin{abstract}
Even though transitivity is a central structural feature of social networks, its influence on epidemic spread on coevolving networks has remained relatively unexplored. Here we introduce and study an adaptive SIS epidemic model wherein the infection and network coevolve with non-trivial probability to close triangles during edge rewiring, leading to substantial reinforcement of network transitivity.  This new model provides a unique opportunity to study the role of transitivity in altering the SIS dynamics on a coevolving network. Using numerical simulations and Approximate Master Equations (AME), we identify and examine a rich set of dynamical features in the new model.  In many cases, the AME including transitivity reinforcement provides accurate predictions of stationary-state disease prevalences and network degree distributions. Furthermore, for some parameter settings, the AME accurately trace the temporal evolution of the system. We show that higher transitivity reinforcement in the model leads to lower levels of infective individuals in the population, when closing a triangle is the dominant rewiring mechanism. These methods and results may be useful in developing ideas and modeling strategies for controlling SIS type epidemics.
\end{abstract}
\maketitle

\section{Introduction}

In recent years, the study of dynamical processes on complex networks has received significant attention in the mathematical modeling of epidemics \cite{keeling2005networks, newman2002spread}.  There exist three distinct approaches to modeling epidemic spread as a dynamical process on networks: in the first, each node is allowed to change its state with no evolution of the underlying network structure through time, while in the second the network structure co-evolves with the state of the nodes \cite{gross2008adaptive, sayama2013modeling}. A third and more complex approach involves combining the inherent (disease-independent) evolution of social networks with the epidemic dynamics \cite{vazquez2007impact, karsai2011small, vestergaard2014memory, holme2014birth}. The second approach is more realistic for modeling epidemic spread in cases where the state of being infected influences the presence of contacts over which the infection can spread. Pragmatically, the details of the network evolution incorporates potentially several social processes. It is a common observation that healthy individuals avoid contact with individuals suffering from an infectious disease. A recent study shows that humans can identify sick individuals through olfactory-visual cues; furthermore, the human immune system appears to use these cues to motivate healthy individuals to avoid contacts with infective individuals \cite{Regenbogen6400}, reducing the risk of contagions and increasing the biological fitness.  Another mechanism that leads to the severing of ties between healthy and infective individuals is the practice of quarantine, often enforced by public health agencies to stop the spread of communicable diseases. Such a preference in attachment between individuals changes the social network and thus influences the spread of the infection. Studying this interplay between the infection spread and the coevolving network structure has the potential to provide new insights into the processes of disease spread.

The features of an underlying social network can govern the propagation of an infectious disease \cite{read2008dynamic, hill2010infectious,shaw2008fluctuating, zanette2008infection, funk2010modelling}, and there is an extensive literature exploring the impact of network structure on the dynamics of epidemic spread on static networks \cite{read2008dynamic, hill2010infectious,shaw2008fluctuating, zanette2008infection, funk2010modelling,moore2000epidemics,miller2009percolation, kuperman2001small, hufnagel2004forecast}. For instance, Ref.~\cite{miller2009percolation} shows that clustering in networks could raise the epidemic threshold and degree correlations can alter the epidemic size. However, results obtained using static network models ignore effects due to the underlying social networks evolving in response to the spread of a disease. Coevolutionary network systems, also called ``adaptive networks'', model the dynamics of node states (e.g., susceptible v.s. infected) together with rules for rewiring network edges in response to the observed states, providing a more general setting to model disease dynamics.  Numerous studies have explored different aspects of the interplay between coevolving networks and epidemic spread \cite{newman2009random, gleeson2009bond}, including extensions to signed networks \cite{saeedian2017epidemic}; nevertheless, a variety of open problems and challenges still exist despite increased understanding of coevolving systems \cite{ebel2002dynamics, holme2006nonequilibrium, scarpino2016effect}.  

A common problem across many coevolving network systems is that they often employ edge rewiring rules that ignore transitivity, a fundamental structural property of most networks, wherein friends of friends are more likely to be friends. For example, the typical random-rewiring rule would randomize any given network and destroy most closed triangles present in the initial conditions; this rule is very unlikely to create new triangles in the network and yields a trivial transitivity level corresponding to that expected under edge independence assumptions. Some recent attempts have been made to explore influences of transitivity in coevolving voter models \cite{malik2013role,malik2016transitivity,thurner2016multiplex}. However, the effect of transitivity on coevolving network epidemics remains unstudied, despite the known importance of transitivity in the static network setting.

In this paper, we introduce a new adaptive SIS model on a coevolving network which incorporates a rewiring rule that reinforces transitivity similar to that employed in \cite{malik2016transitivity,thurner2016multiplex}. The most straightforward strategy to deal with an infectious disease is to sever all the links between susceptible and infected individuals. However, humans are social animals and need a certain amount of social relationships to function. Therefore, only discarding the infected-susceptible links without compensation is not pragmatic in all settings. Under the mechanism considered here, susceptible individuals break links to their infected neighbors, rewiring those links to either a random individual in the network or a neighbor's neighbor. The latter rewiring will lead to reinforcement of transitivity (increased clustering coefficient) in the system.  While transitivity is a critical property of social networks, studies of epidemic dynamics on coevolving networks have mostly neglected this property until now. Our  model thus provides an opportunity to study the role of transitivity on epidemic dynamics in coevolving network systems. This rewiring rule includes a parameter that governs the preferential rewiring to close triangles. We study the impact of this transitivity reinforcement on the spread of the infection and on the details of the network using numerical simulations of the process as well as an Approximate Master Equation (AME) framework (see \cite{gleeson2011high,gleeson2013binary}). In Sec.~\RN{2}, we describe our model and in Sec.~\RN{3} we present results obtained through numerical simulations, highlighting the new dynamical features observed in the model. In Sec.~\RN{4},  we present a derivation of our semi-analytical method, confirming results obtained from numerical simulations. Secs.~\RN{3} and Sec.~\RN{4} also include identification and analysis of bifurcations in the system dynamics. Finally, we conclude in Sec.~\RN{5} with further discussion.

\section{Model}

We employ the classical setting of the susceptible-infected-susceptible (SIS) model \cite{brauer2001mathematical, gross2006epidemic}. We consider a network with $N$ nodes and $M$ undirected edges (both constant), with $M=\langle k \rangle N/2$ where $\langle k \rangle$ is the average degree.  At a given time, each node is either in a susceptible (S) or infected (I) state.  Infected individuals infect each of their susceptible neighbors at rate $\beta$ while recovering (change to susceptible) at rate $\alpha$. Meanwhile, each susceptible individual breaks each edge it has with an infected neighbor at rate $\gamma$, rewiring this link to a susceptible node. 

Generalizing previously studied models, the new link created during rewiring is made with probability $\eta$ to a susceptible node at distance two---that is, a neighbor of a neighbor to whom the given node is not already connected. Otherwise (i.e., with probability $1-\eta$) the rewiring is made to another susceptible node selected uniformly at random from the network outside its immediate neighborhood [that is, self-loops and repeated links (multiedges) are prohibited]. Under this latter, uniform rewiring, there remains non-zero probability of randomly picking a neighbor of a neighbor; but this probability becomes vanishingly small for larger networks. In the event that there is no available susceptible node to rewire to (e.g., there are no distance-two susceptible nodes), the rewiring attempt fails and the original link to the infected neighbor remains. Importantly, rewiring to a neighbors' neighbor closes a triangle between the three nodes, directly reinforcing transitivity, whereas rewiring uniformly at random tends to decrease transitivity.

\section{Numerical Simulation}
Unless stated otherwise, our numerical results of the adaptive SIS system presented here are for networks with $N = 25,000$ nodes and $M = 25,000$ edges (that is, mean degree $\langle k \rangle = 2$); and 1,000 simulations of the process are produced for each configuration of parameters.  We performed Monte Carlo simulations with the help of \texttt{largenet}, a C++ library that has been widely used for simulations of large adaptive networks \cite{zschaler2012largenet2}. We used \texttt{largenet} with fixed step size $1/N$, asynchronously updating one randomly-selected node at a time.

Following Ref.~\cite{marceau2010adaptive}, we consider the following three initial degree distributions: the Poisson distribution
\begin{equation}
p_{k}^{P} = \frac{\langle k \rangle ^{k}e^{-\langle k \rangle }}{k!}\,, \label{eq:pkp}
\end{equation}
approximating the distribution of an Erd\H{o}s-R\'{e}nyi model for large $N$; a truncated power law distribution
\begin{equation}
p^{TPL}_{k} = \left\{
     \begin{array}{cc}
       \frac{1}{C} k^{-\tau}   &  ~ 0<k \leqslant k_{c}\\
       0 & ~ k>k_{c}\,,
     \end{array}
   \right.
   \label{eq:tpl}
   \end{equation}
with $\tau = 2.161$ and $k_{c} = 20$ in order to make the mean degree $\langle k \rangle = 2$ (we also study the effect of different cut-off degrees in Appendix A.1); and the degree-regular distribution
\begin{equation} 
p_{k}^{DR} = \delta_{k,k_{0}}\,, \label{eq:drd} 
\end{equation}
where $\delta$ is the Kronecker delta indicating every node has the same initial degree $k_{0}=2$.

We observe that at large enough time $t$, the system approaches a statistically stationary state with features of interest such as the clustering coefficient and disease prevalence fluctuating around mean values. For the purposes of identifying this state numerically, we determine the stationary state to be reached when the level of disease prevalence $I$ (expressed as a fraction of nodes) between two consecutive integer times differs by at most ${10}^{-5}$. Under this definition, we find that in practice networks with the above three initial degree distributions reach their stationary states before $t = 5,000$ for $\eta < 1$. In contrast, when $\eta = 1$ we do not observe convergence to stationary states, even after a long time; rather, we appear to have a long, progressively slowing decay of the disease prevalence (as we will see below in Fig.~\ref{ER_dynamics}).

\begin{figure}
    \centering
    \includegraphics[width=\columnwidth]{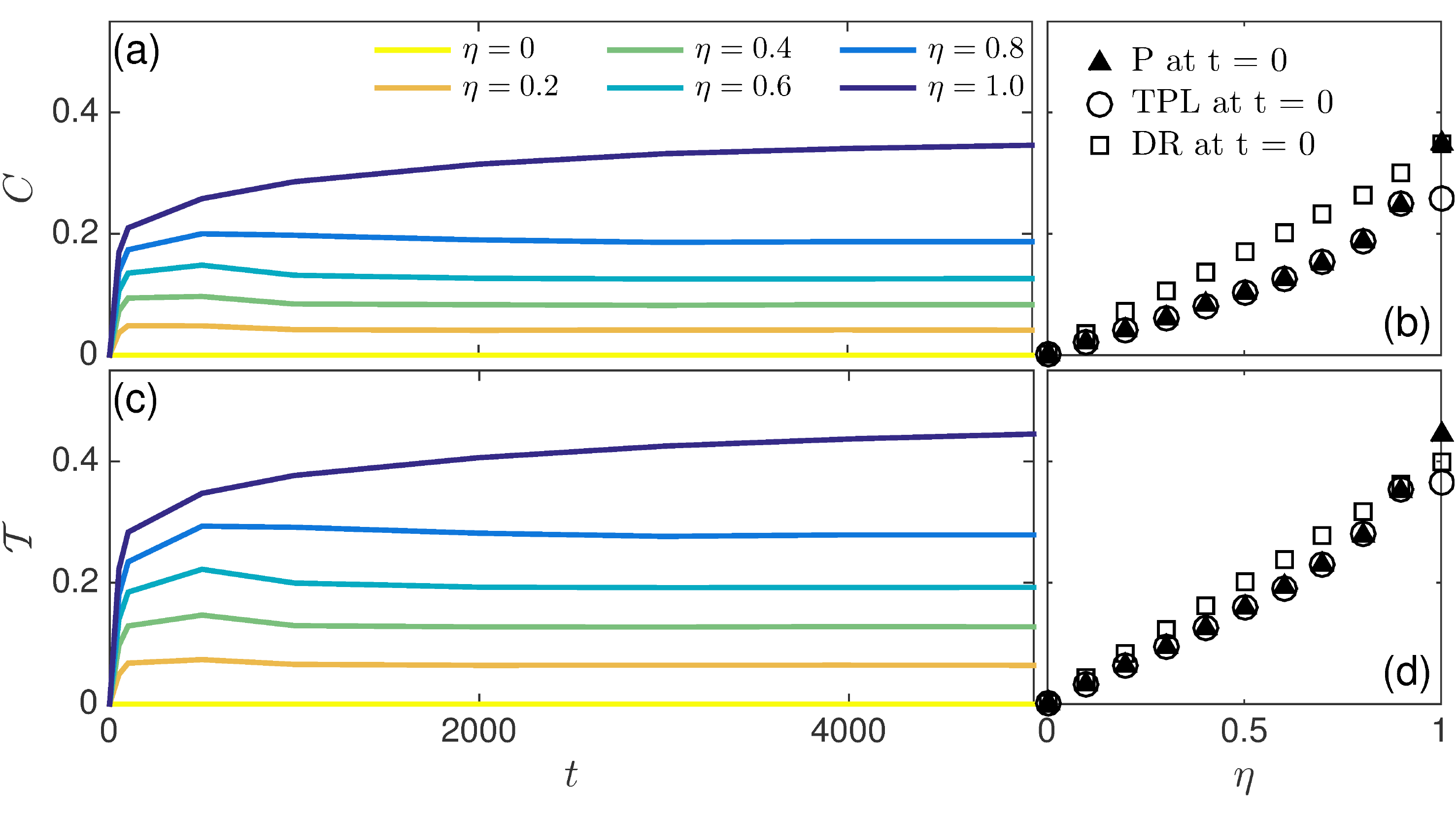}
    \caption{The level of transitivity varies with time $t$ and the probability $\eta$ of rewiring steps made to susceptible neighbors of neighbors. Panels (a) and (c) show the average local clustering coefficient $C$ and transitivity $\mathcal{T}$, respectively, versus time $t$ for simulations with initially Poisson degree distribution with mean degree $\langle k \rangle = 2$. Other parameters of the system include $\beta = 0.04$, $\gamma = 0.04$, $\alpha = 0.005$, and $\epsilon = 0.1$. Panels (b) and (d) show $C$ and $\mathcal{T}$, respectively, at $t=5,000$ versus $\eta$ for networks with the same mean degree $\langle k \rangle = 2$ but different initial degree distributions [$p^{P}_{k}$ (Poisson): triangle, $p^{TPL}_{k}$ (truncated power law): circle, and $p^{DR}_{k}$ (degree regular): square]. Plotted data corresponds to mean values computed over 30 Monte Carlo simulations. In (a) and (c), all cases except $\eta=1$ appear to reach stationarity by time $t = 5,000$.}
    \label{CT_plot}
\end{figure}

\begin{figure}
 \centering
    \includegraphics[width=\columnwidth]{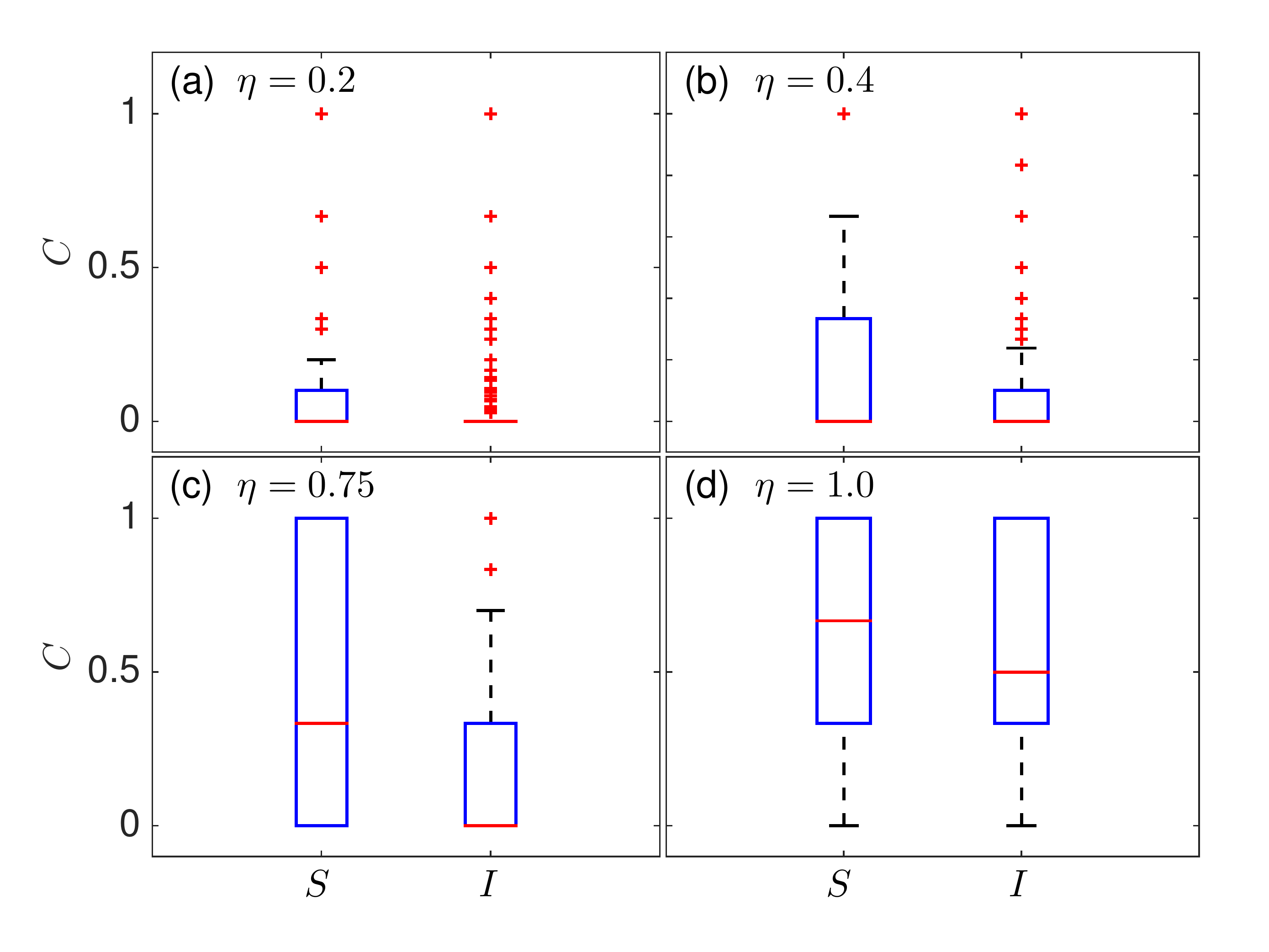}
    \caption{Local clustering coefficients $C$ grouped by susceptible nodes ($S$) and infected nodes ($I$) at time $t = 5,000$ for (a) $\eta = 0.2$, (b) $\eta = 0.4$, (c) $\eta = 0.8$, and (d) $\eta = 1$. The initial degree distribution is Poisson. The other parameters of the system are $\beta = 0.04$, $\gamma = 0.04$, $\alpha = 0.005$, $\epsilon = 0.1$, and $\langle k \rangle = 2$. At time $t = 5,000$, all cases except for $\eta = 1$ reach their stationary states. Each boxplot is obtained from a single Monte Carlo simulation of network size $N = 25,000$. The central red mark indicates the median, the bottom and top edges of the box indicate the 25th and 75th percentiles, respectively, whiskers extend to the most extreme data points not considered as outliers, and the outliers are plotted individually (red ``$+$'').}
    \label{individual_C}
\end{figure}

To highlight the main differences between our model and previous work without reinforced transitivity \cite{marceau2010adaptive}, we first confirm that our model leads to non-trivial transitivity. We explore two closely related measures of  transitivity here. The first measure is the average local clustering coefficient $C=\displaystyle\frac{1}{N} \sum_{i=1}^{N} C_i $;  where the local clustering coefficient  $C_i$ of vertex $i$ is  $C_i={T_i}/{\tau_i}$, $T_i$  and  $\tau_i$ are the number of closed triplets (triangles) and the total number of triplets centered at vertex $i$, respectively \cite{watts1998collective}.  The second measure is the  transitivity $\mathcal{T}$ (often called  global clustering coefficient) \cite{wasserman1994social},  defined as: $\displaystyle\mathcal{T}={3T}/{\tau}$, where $T$ and $\tau$  are, respectively, the total number of closed triangles and  the number of triplets in the network. Fig.~\ref{CT_plot} panels (a) and (c) show the temporal evolution of  the average local clustering coefficient $C$ and the transitivity $\mathcal{T}$ in our model, with varying $\eta$ from $0$ to $1$ in a step of $0.2$, keeping all other parameters fixed and choosing the initial degree distribution to be Poisson. When $\eta = 0$,  there is no reinforcement of transitivity, i.e., $C \approx 0$ ($\mathcal{T} \approx 0$) for all time and we are in the same regime studied in Ref.~\cite{marceau2010adaptive}. For nonzero $\eta$, the clustering coefficients rapidly increase and then slowly converge to nonzero values, except in the case $\eta = 1$ where we do not observe this convergence, and  $C$ (and $\mathcal{T}$) appears to continue to grow with a diminishing rate.  We treat $\eta = 1$ as a special case in our model and will study this case separately later. We also explored the influence of initial degree distributions on the final clustering coefficient. In Fig.~\ref{CT_plot}(b) and (d), we observe that networks with the regular degree  distribution initially achieve higher clustering coefficients compared to networks with the other two initial distributions. Moreover, networks with initial Poisson and truncated power law degree distributions show indistinguishable final clustering coefficients for $\eta < 1$. Lastly, in Fig.~\ref{individual_C}, we examine the local clustering coefficient $C$ grouped by susceptible nodes ($S$) and infected nodes ($I$) at time $t = 5,000$ for networks with Poisson degree distributions initially and with different values of $\eta$. As $\eta$ increases, susceptible nodes typically achieve higher values of local clustering coefficient $C$ compared to the infected nodes. 

\begin{figure}
\centering
\includegraphics[width = \columnwidth]{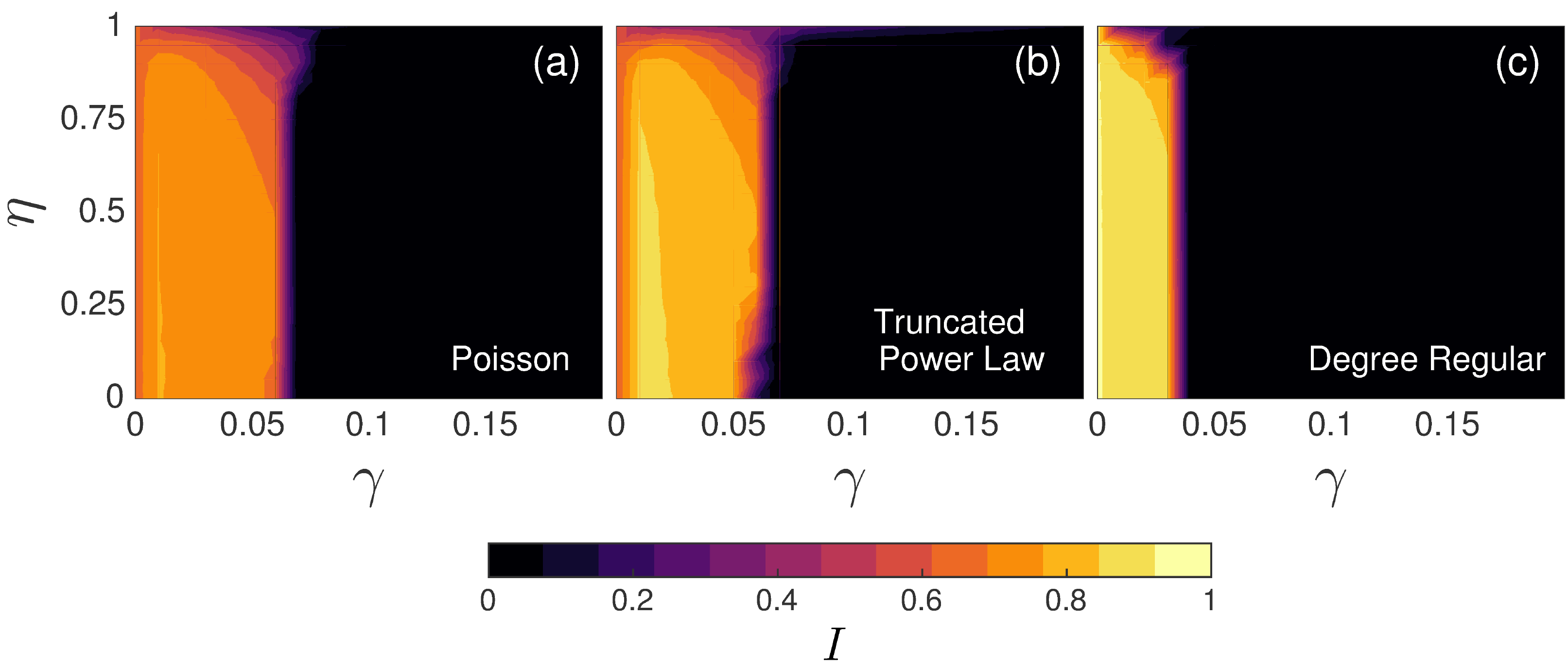}
\caption{Phase diagrams in parameter space $(\gamma,\eta)$ for the observed disease prevalence, $I$, at time $t=10,000$ on networks with initial degree distributions that are (a) Poisson, (b) truncated power law, and (c) degree regular. We here fix $\beta = 0.04$, $\alpha = 0.005$ and $\epsilon = 0.1$. At the selected time $t = 10,000$ all cases except $\eta=1$ appear to have converged to their stationary states. Results are averaged over 30 simulations at each set of parameters. These diagrams are generated through bilinear interpolation from results on a regular grid, leading to some apparent grid artifacts.}
\label{phase_plot}
\end{figure}

Given that our model demonstrates non-trivial transitivity, we now explore how the reinforced transitivity influences the evolution of disease prevalence, $I$. In Fig.~\ref{phase_plot} we plot the observed disease prevalence at long times in the $(\gamma,\eta)$ parameter space, fixing the infection rate $\beta = 0.04$ and recovery rate $\alpha=0.005$ as in Ref.~\cite{marceau2010adaptive}. The phase diagram in Fig.~\ref{phase_plot} indicates that for large enough $\gamma$ the system eventually becomes disease free, presumably because susceptible nodes can easily rid themselves of infected neighbors by rewiring these edges, thus effectively quarantining the infection. When $\gamma = 0$, there is no rewiring process, the network does not change, and hence $\eta$ has no effect. A more interesting behavior occurs for $\gamma \lesssim 0.05$, with large values of disease prevalence that depend on $\eta$ in a complex manner. Since this parameter regime is most interesting for our present study, in the rest of our simulations we fix $\gamma = 0.04$ and vary $\eta$ to investigate the influence of transitivity reinforcement.

\begin{figure}[!ht]
 \centering
    \includegraphics[width=\columnwidth]{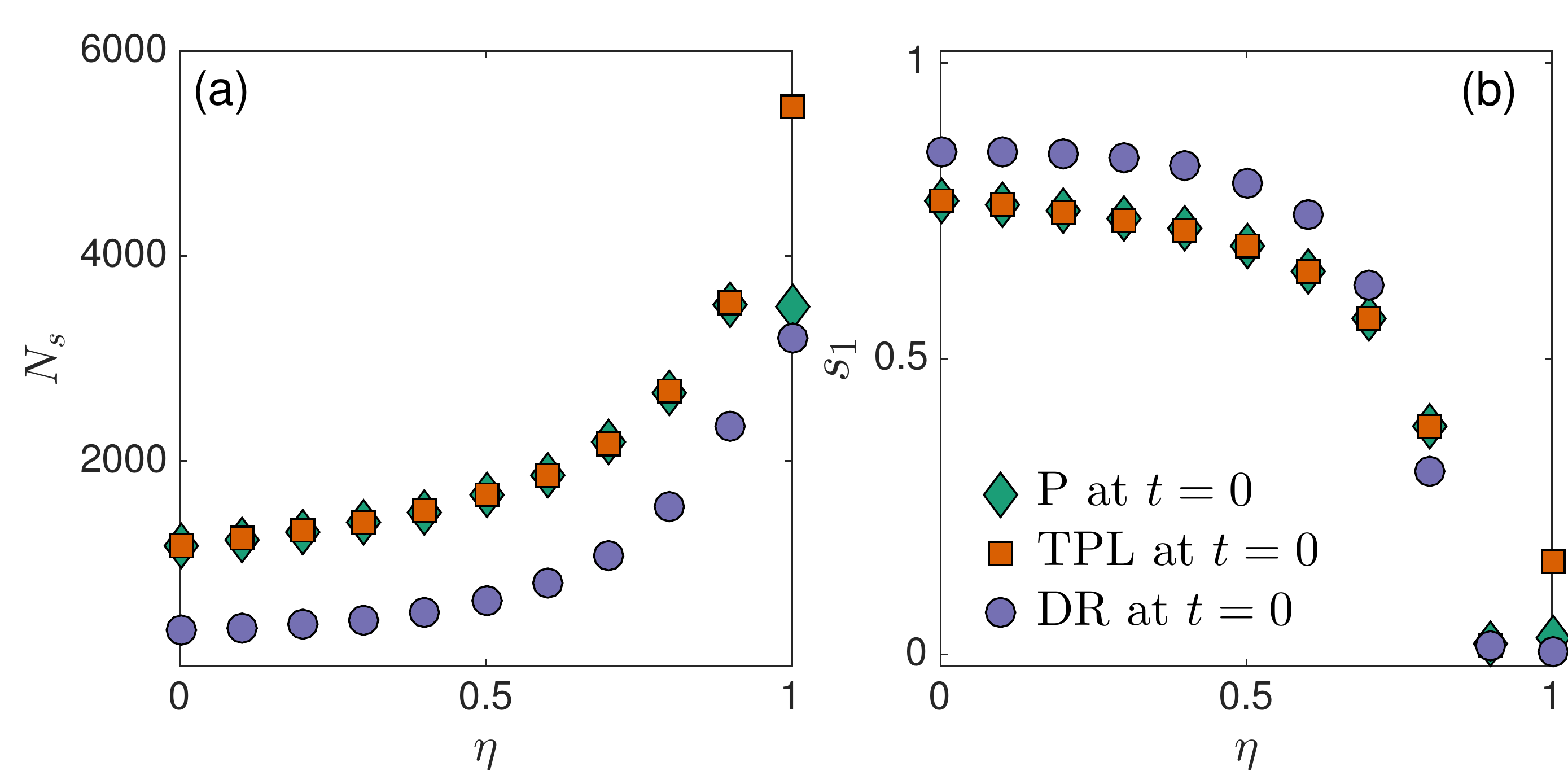}
    \caption{(a) Number of components $N_s$ and (b) Giant component fraction $s_1$ versus $\eta$ at $t = 5,000$. The initial degree distribution of networks are Poisson (P), truncated power law (TPL), and degree regular (DR) with mean degree $\langle k\rangle = 2$. Other parameters in the system are $\gamma = 0.04$, $\beta = 0.04$, $\alpha = 0.005$, and $\epsilon = 0.1$. Every point in the figure corresponds to the mean value computed over 30 simulations. In these simulations, the mean number of components at $t = 0$ for Poisson (P), truncated power law (TPL), and degree regular (DR) are $666.17$, $3348.27$, and $6.03$, respectively. And the mean giant component fraction at $t = 0$ for Poisson (P), truncated power law (TPL), and degree regular (DR) are $0.7971$, $0.6828$, and $0.7452$, respectively. All cases except $\eta = 1$ reach stationarity by time $t = 5,000$.}
    \label{giant_component}
\end{figure}

Another critical phenomenon that typically occurs in coevolving network models is the fission of the underlying network as the system evolves.  In Fig.~\ref{giant_component}, we explored the number of connected components ($N_s$) and the fraction of nodes in the giant component ($s_1$) at $t = 5,000$ for the three initial settings of the degree distribution.  We fixed the mean degree $\langle k\rangle = 2$, $\gamma = 0.04$, $\beta = 0.04$, $\alpha = 0.005$, and $\epsilon = 0.1$ and then computed $N_s$ and $s_1$ at $t = 5,000$. Varying $\eta$ from 0 to 1, we observe the increase in $N_s$ and decrease in $s_1$, indicating that higher $\eta$ leads to more fragmented networks in the stationary states.  Interestingly, the results corresponding to the Poisson (P) and the truncated power law (TPL) degree distributions still overlap except when $\eta = 1$.

\begin{figure*}
\centering
\includegraphics[width = 1.75\columnwidth]{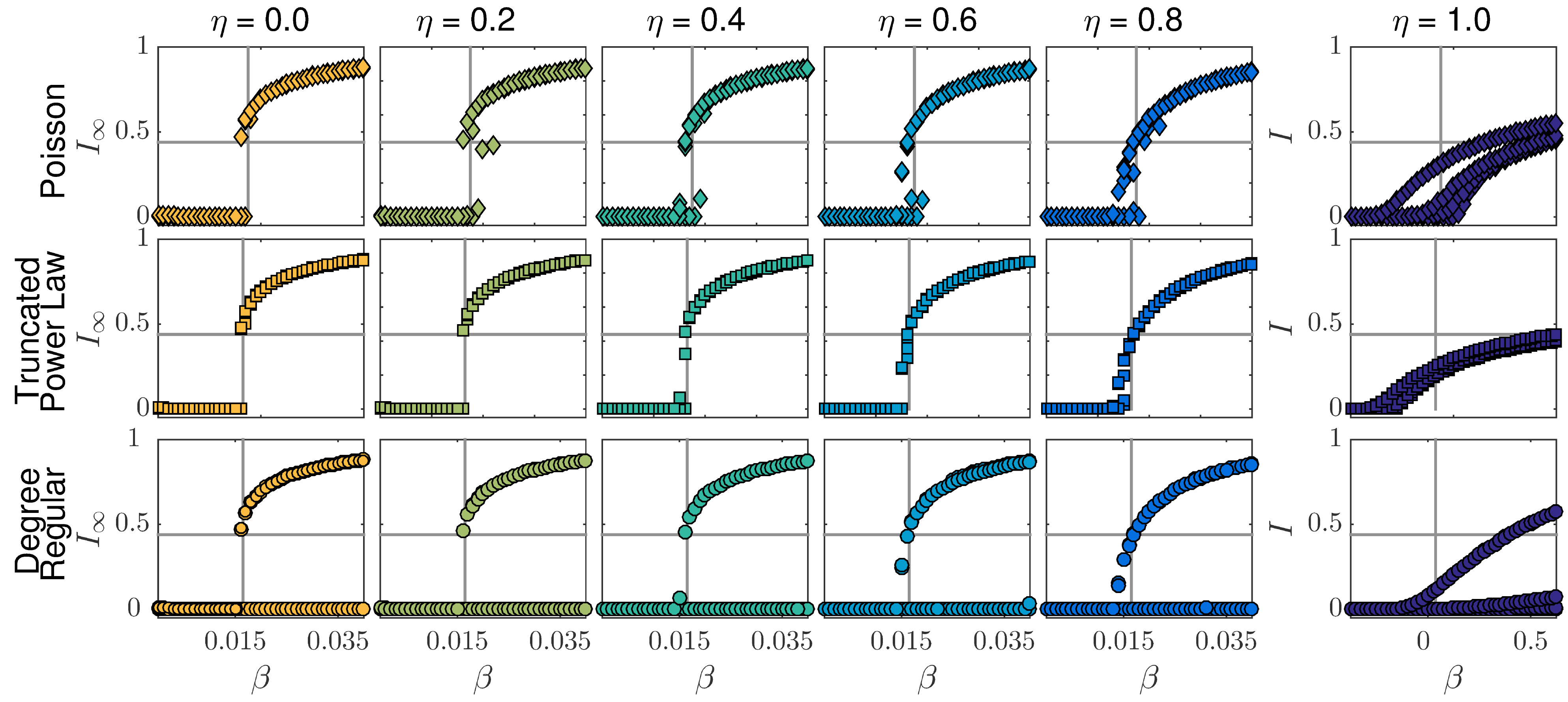}
\caption{Bifurcation diagrams of the disease prevalence in the stationary state  $I_{\infty}$ versus $\beta$ at different values of $\eta$. Each row of plots corresponds to a specific degree distribution: the first row is Poisson, the second row is truncated power law, and the third row is degree regular. Each column of plots corresponds to a particular value of $\eta$, as indicated at the top of the column.  Every point represents a mean of 30 realizations of the system at $t= 10,000$. The last column belongs to the $\eta=1$ case, noting that the $y$-axis is labeled $I$ (cf.\ $I_{\infty}$) because the simulations have not reached stationarity. Other parameters of the system are $\gamma = 0.02$, and $\alpha = 0.005$. (We also study the effect of different $\epsilon$ in initially Poisson networks, see Appendix A.2.)}
\label{bifurcation_1}
\end{figure*}

To further explore the dynamics of the system, we plot bifurcation diagrams in Fig.~\ref{bifurcation_1}.  The rows in Fig.~\ref{bifurcation_1} correspond to different initial degree distributions with columns for different values of $\eta$.  When $\eta < 1$, for all three initial distributions there exists a small bistable region near $\beta \doteq 0.16$. The sudden jump of $I_\infty$ around the critical $\beta$ when $\eta = 0$  (the first column of plots) implies that the transition might be discontinuous when $\eta = 0$. This discontinuity appears to shrink as  $\eta$ increases. For $\eta = 0.8$, the transition appears to be continuous. While we do not conclusively confirm the nature of the transitions here, we will investigate the transitions in more detail in Section V. We emphasize the separation of figures corresponding to $\eta=1$, with y-axis labeled as $I$ and not $I_\infty$, highlighting the fact that for $\eta=1$ the system does not reach a stationary state in the observed time. 

\section{Approximate Master Equations}

\begin{figure}
\includegraphics[width = 0.7\columnwidth]{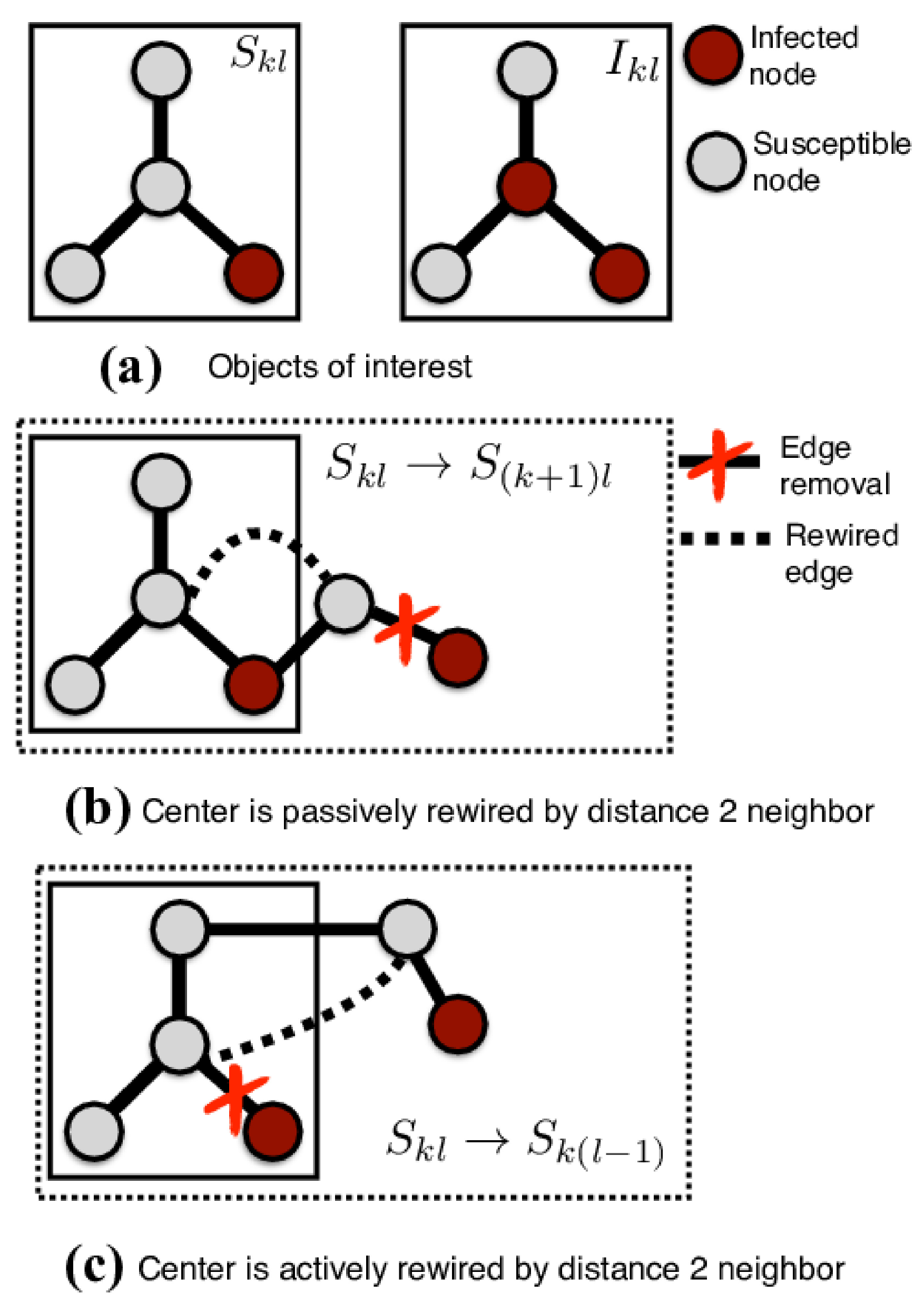}
\caption{Illustration of an S node rewiring to a neighbors' neighbor. (a) Before the rewiring, the center nodes in the diagram are of class $S_{kl}$ and $I_{kl}$, respectively. Consider the two S nodes involved when an $SI$ edge is rewired to a distance-two neighbor: the active node doing the rewiring and the passive node receiving the newly-rewired edge. (b) This diagram represents the case when the center is \emph{passively} rewired by the action of a distance-two neighbor, with the center node moving to class $S_{(k+1)l}$. (c) This diagram represents the case when the center \emph{actively} rewires to a distance-two neighbor, and the center node becomes class $S_{k(l-1)}$. The other processes of the adaptive SIS system can be similarly diagrammed.}
\label{illustration}
\end{figure}

We further study this system with approximate semi-analytical techniques, defining appropriate systems of ordinary differential equations and then solving those systems numerically. The moment-closure frameworks of Pair Approximation (PA) and Approximate Master Equations (AME) have both been used effectively in similar settings \cite{durrett2012graph, gleeson2011high, keeling2005networks}. In PA, the different counts of contact pairs are used to approximate the density of triplets, thus closing the system of equations \cite{gleeson2011high}. In contrast, the AME framework considers the populations of nodes according to their state, degree, and the states of their immediate neighbors, approximating other quantities from these populations. AME is an annealed mean field approximation that deterministically approximates stochastic systems. We note in particular that with the additional variables and corresponding differential equations in AME, the triplet counts are precisely accounted for (cf.\ the closure approximations necessary to obtain triplet counts in PA) and that AME typically provides a more accurate approximation of such dynamics and can be highly stable around the critical point of the dynamics \cite{durrett2012graph,gleeson2011high,keeling2005networks,zhou2013link,marceau2010adaptive} (see also \cite{gleeson2011high, durrett2012graph, lee2017evol} for comparisons between AME and PA).  We note that Ref.~\cite{zhou2013link} further extended this approach to classify links according to the states, numbers of neighbors, and numbers of infected neighbors at both ends of the links. This link-based approach can further improve on the accuracy of the AME method; however, it increases the system size from $O(k^{2}_{max})$ to $O(k^{4}_{max})$ equations, where $k_{max}$ is the maximum degree. As such, we restrict our attention here to the AME method.

We employ an approach similar to that used in Ref.~\cite{marceau2010adaptive}, extending the equations to include the effect of transitivity reinforcement. Let $S_{kl}(t)$ and $I_{kl}(t)$ be the fraction of susceptible and infected nodes, respectively, of degree $k$ with $l$ infected neighbors at time $t$ [see diagram in Fig.~\ref{illustration}(a)]. Following the notation in Ref.~\cite{marceau2010adaptive}, we define the zeroth-order moments of the $S_{kl}(t)$ and $I_{kl}(t)$ distributions by 
\begin{equation*}
S \equiv \sum_{kl}S_{kl} \text{ and } I \equiv \sum_{kl}I_{kl}\,; 
\end{equation*}
the first order moments by  \begin{equation*}
\begin{split}
& S_{S} \equiv \sum_{kl}(k-l)S_{kl}\,, \text{ } S_{I} \equiv \sum_{kl}lS_{kl}\,, \\
& I_{S} \equiv \sum_{kl}(k-l)I_{kl}\,, \text{ and } I_{I} \equiv \sum_{kl}lI_{kl}\,; 
\end{split}
\end{equation*}
and the second order moments by  \begin{equation*}
S_{SI} \equiv \sum_{kl}(k-l)lS_{kl}\,, \text{ } S_{II} \equiv \sum_{kl}l(l-1)S_{kl}\,, \text{ and so on.}
\end{equation*}
It is worth noting that while node states and network topologies coevolve, some quantities are conserved. For example, $S + I = 1$, since the number of nodes is fixed. Similarly, $S_S + S_I + I_S + I_I = \langle k \rangle$, because of the conservation of edges. We then have the following ODE governing the time evolution of the $S_{kl}$ compartment:
\begin{equation}
\begin{split}
\frac{dS_{kl}}{dt} & = \alpha I_{kl} - \beta l S_{kl} +\alpha \big[ (l+1)S_{k(l+1)} - lS_{kl} \big]\\
& + \beta \frac{S_{SI}}{S_{S}} \big[ (k-l+1)S_{k(l-1)} - (k-l)S_{kl}\big]\\
& + \gamma \big[(l+1)S_{k(l+1)} - l S_{kl}\big]\\
& + \gamma \eta \bigg\{ \big[  \frac{l}{k-1} \frac{I_{S}}{\frac{1}{2}I_{I}+I_{S}} \frac{S_{I}}{S}\\
&    ~~~~~~ + \frac{k-l-1}{k-1} \frac{\frac{1}{2}S_{S}}{\frac{1}{2}S_{S}+S_{I}} \frac{S_{I}}{S} \big] S_{(k-1)l}\\ 
& - \big[\frac{l}{k} \frac{I_{S}}{\frac{1}{2}I_{I}+I_{S}} \frac{S_{I}}{S} + \frac{k-l}{k} \frac{\frac{1}{2}S_{S}}{\frac{1}{2}S_{S}+S_{I}} \frac{S_{I}}{S} \big] S_{kl} \bigg\}\\
& + \gamma (1-\eta) \frac{S_{I}}{S} \big[S_{(k-1)l} - S_{kl} \big].
\end{split}
\label{eq:AME-S}
\end{equation}
Similarly the ODE for the $I_{kl}$ compartment is
\begin{equation}
\begin{split}
\frac{dI_{kl}}{dt} & = -\alpha I_{kl} + \beta l S_{kl} +\alpha \big[ (l+1)I_{k(l+1)} - lI_{kl} \big]\\
& + \beta \big(1+ \frac{S_{II}}{S_{I}} \big) \big[ (k-l+1)I_{k(l-1)} - (k-l)I_{kl}\big]\\
& + \gamma \big[(k-l+1)I_{(k+1)l} - (k-l) I_{kl}\big]. 
\end{split}
\label{eq:AME-I}
\end{equation}

\begin{figure}
\centering
\includegraphics[width = \columnwidth]{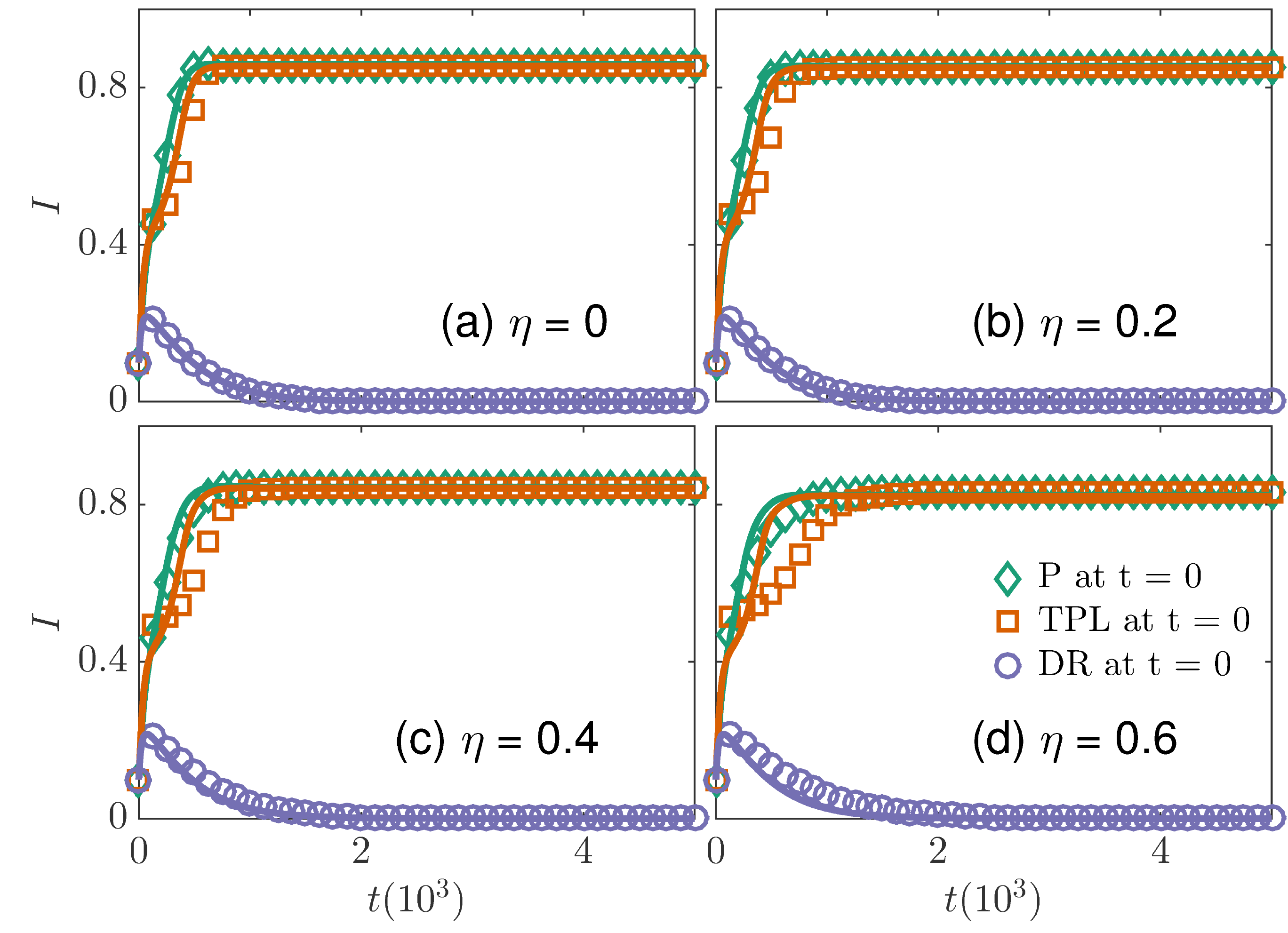}
\caption{Disease prevalence $I$ at time $t$.  Initial degree distributions are Poisson (diamonds, $P$ at $t=0$, see Eq.~\ref{eq:pkp}), truncated power law (squares, $TPL$ at $t=0$, see Eq.~\ref{eq:tpl}) and degree regular (circles, $DR$ at $t=0$, see Eq.~\ref{eq:drd}). Markers correspond to means computed over 1,000 simulations and lines are the semi-analytical AME results. The parameters used are $\beta = 0.04$, $\gamma = 0.04$, $\alpha = 0.005$, $\epsilon = 0.1$ and  $\langle k \rangle = 2$, with (a) $\eta = 0$, (b) $\eta = 0.2$, (c) $\eta = 0.4$, (d) $\eta = 0.6$.
}
\label{disease_prevalence}
\end{figure}

To describe the AME derivation, we will focus on explaining the $S_{kl}$ equation, as the derivation of corresponding terms in the $I_{kl}$ equation involves similar arguments. The terms pre-multiplied by $\alpha$ and $\beta$ in the first two lines of equation (\ref{eq:AME-S}) account for the ``center" node and its neighbors recovering from and becoming infected. The third line, pre-multiplied by $\gamma$, describes the ``center" class in state $S$ dismissing one of its state $I$ neighbors and rewiring to a random node with state $S$. For more detailed discussion on the derivation of these terms, we refer to Ref.~\cite{marceau2010adaptive}. 

The next three lines of equation (\ref{eq:AME-S}), pre-multiplied by $\gamma \eta$ are new to the present work, accounting for a ``center" class with state $S$ and its distance two neighbors with state $S$, breaking links to one of their $I$ neighbors and rewiring to the center node (see the diagrams in Fig.~\ref{illustration}). To better understand the derivation of these lines, we note that only $S$ nodes can rewire and be rewired to in this model (whereas $I$ neighbors are only dropped by the action of $S$ nodes).  Supposing the center node is of class $S_{(k-1)l}$, then it has $\frac{l}{k-1}$ proportion of its neighbors infected neighbors, and we describe the number of $SI$ edges among them as $\frac{I_{S}}{\frac{1}{2}I_{I}+I_{S}} \cdot \frac{S_{I}}{S}$, making assumptions about independence to close the equations with this approximation. Similarly, the same center node class $S_{(k-1)l}$ has $\frac{k-l-1}{k-1}$ proportion of its neighbors susceptible, with the number of $SI$ edges among them described by $\frac{\frac{1}{2}S_{S}}{\frac{1}{2}S_{S}+S_{I}} \cdot \frac{S_{I}}{S}$. Therefore, the rate associated with the transition $S_{(k-1)l}$ to $S_{kl}$ is  $\frac{l}{k-1} \frac{I_{S}}{\frac{1}{2}I_{I}+I_{S}}  \frac{S_{I}}{S} + \frac{k-l-1}{k-1}\frac{\frac{1}{2}S_{S}}{\frac{1}{2}S_{S}+S_{I}}  \frac{S_{I}}{S}$. Following the same arguments, we find that the corresponding rate for a node leaving the $S_{kl}$ class is  $\frac{l}{k} \frac{I_{S}}{\frac{1}{2}I_{I}+I_{S}}  \frac{S_{I}}{S} + \frac{k-l}{k}\frac{\frac{1}{2}S_{S}}{\frac{1}{2}S_{S}+S_{I}}  \frac{S_{I}}{S}$. 

The last line of equation (\ref{eq:AME-S}), pre-multiplied by $\gamma (1 - \eta)$, is the effect of a ``center" class with state $S$ rewired to a random node in the network with state $S$. For the $I_{kl}$ compartment in equation (\ref{eq:AME-I}), the ``center" class can only lose edges due to rewiring and the value of $\eta$ does not change the rate at which edges to $I$ nodes are rewired. Therefore, the equation lacks terms pre-multiplied by $\gamma \eta$ and $\gamma (1 - \eta)$. The resulting system of coupled ordinary differential equations contains $2(k_{max} + 1)^2$ equations, where $k_{max}$ is the maximum degree. 

\setstcolor{blue}

\begin{figure}
 \centering
    \includegraphics[width=\columnwidth]{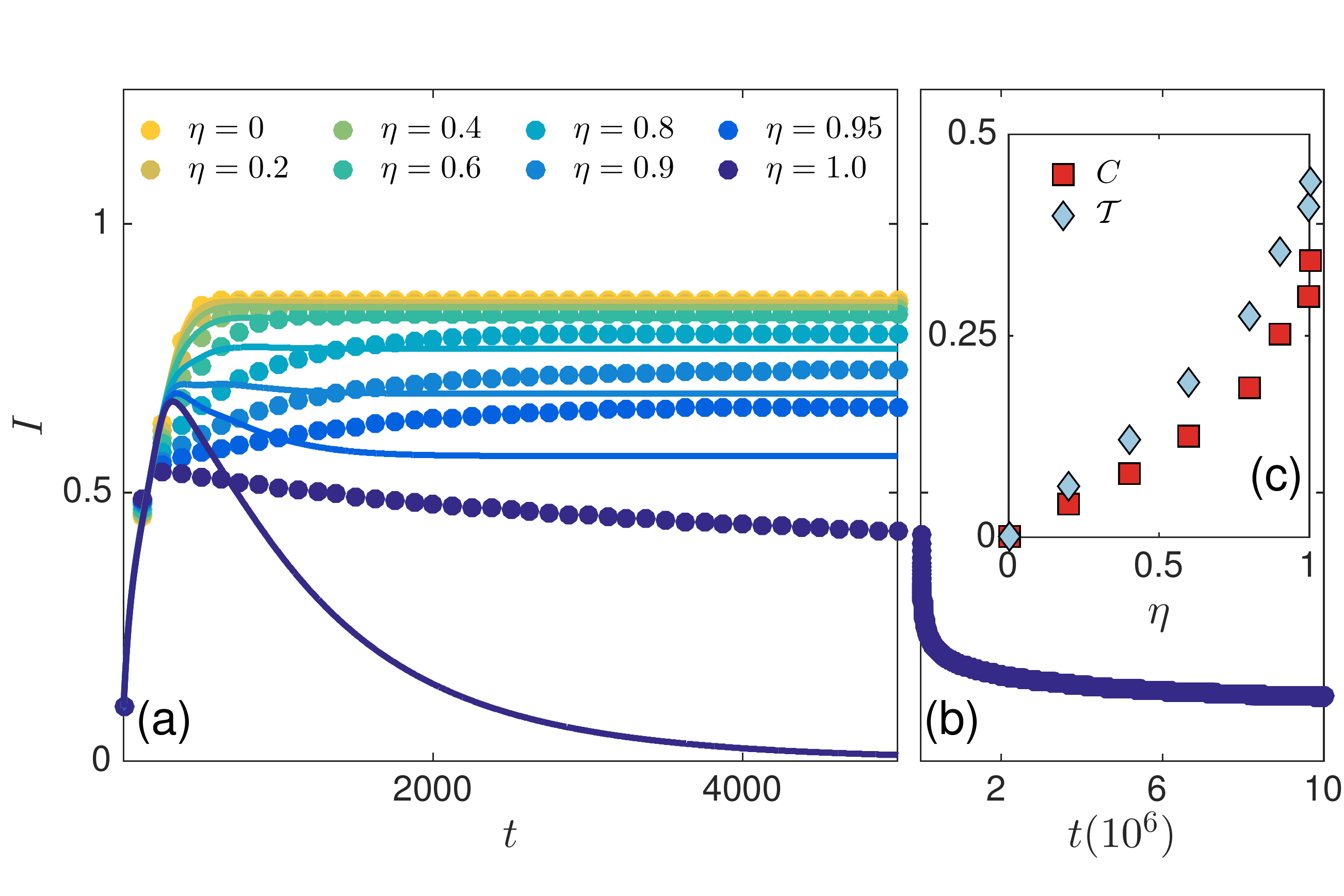}
    \caption{Disease prevalence $I$ at time $t$. The initial degree distribution is Poisson. Different values of $\eta$ are indicated by color. The other parameters of the system are $\beta = 0.04$, $\gamma = 0.04$, $\alpha = 0.005$, $\epsilon = 0.1$, and $\langle k \rangle = 2$. Dots correspond to means computed over 1,000 simulations. Lines are AME results. Observe that for $\eta=1$ the simulations exhibit slow convergence in panel (b) and AME fails to capture the correct time scale for this convergence.  The corresponding clustering coefficient $C$ and transitivity $\mathcal{T}$ at $t = 5,000$ for different $\eta $ are plotted in panel (c).}
    \label{ER_dynamics}
\end{figure}

\begin{figure}
    \hspace*{-0.4cm}\includegraphics[width=1.1\columnwidth]{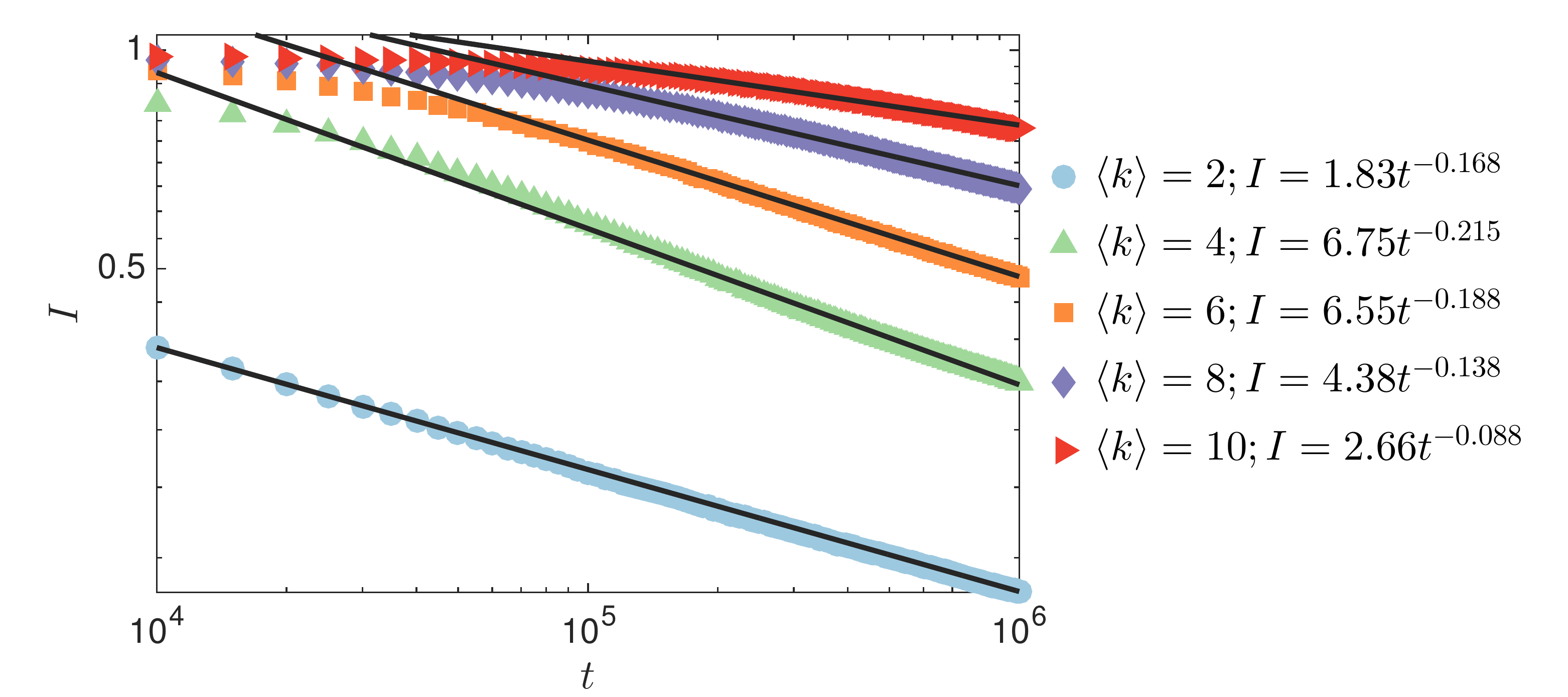}
    \caption{The scaling of disease prevalence $I$ with time when $\eta = 1$. As the average degree $\langle k \rangle$ becomes greater than $4$, the scaling exponent appears to decrease in magnitude, suggesting that this phenomenon may depend on the level of sparsity of the underlying network. The parameters here are $\beta = 0.04$, $\gamma = 0.04$, $\alpha = 0.005$,  $\epsilon = 0.1$ and $N = 25,000$. Markers correspond to means computed over 30 simulations.}
    \label{eta_1}
\end{figure}

To determine the initial conditions for the above equations, initially an $\epsilon$ fraction of randomly chosen nodes are infected, independent of node degree.  This gives us the following set of initial conditions: \begin{align}
\begin{split}
S_{kl}(0) &= (1-\epsilon) p_{k}(0) \binom{k}{l}\epsilon^{l}(1-\epsilon)^{k-l} \\ 
I_{kl}(0) &= \epsilon p_{k}(0) \binom{k}{l}\epsilon^{l}(1-\epsilon)^{k-l},
\end{split}
\end{align}
where $p_{k}(0)$ is the degree distribution at $t = 0$. We proceed to numerically solve the AME system of equations \ref{eq:AME-S} and \ref{eq:AME-I}. In the previous section, we presented an overview of the general behavior of the model via explicit simulations of the model. We now carry out a detailed comparison between the simulations and AME results, confirming the accuracy of AME and furthering our understanding of the adaptive SIS dynamics with reinforced transitivity.

\section{Exploring the system through simulations and AME}

First we plot the disease prevalence $I$ against time $t$ with different $\eta$'s and different initial degree distributions in Fig.~\ref{disease_prevalence}. Note that $\eta = 0$ [Fig.~\ref{disease_prevalence}(a)] is the case where our model reduces to that studied in Ref.~\cite{marceau2010adaptive}.  Networks with different initial degree distributions converge to different stationary disease prevalences. In the Poisson and the truncated power law cases, the disease prevalence $I$ approaches a nonzero value, i.e., the system attains an \emph{endemic} state. However, in the degree-regular case, $I$ gradually approaches $0$, i.e., the system approaches a \emph{disease-free} state. The thick lines in Fig.~\ref{disease_prevalence} represent results derived from the AME approach, while symbols are simulation results.  

Most importantly, AME captures the stationary disease prevalence in all the cases shown in the Figure. Second, for small $\eta$, AME approximates the temporal evolution of the disease prevalence with a high level of accuracy, even in the initial transient states, and correctly predicts the direction of the changes with $\eta$. However, the differences between simulations and AME increase at larger $\eta$ and are particularly obvious for the truncated power law case, but AME still captures the overall shape of the temporal evolution of disease prevalence. (To further highlight these elements, in Fig.~\ref{new_measurements} in Appendix A.3 we provide a zoomed view for the transient dynamics $t \in [0,1000]$ for some of the panels of Fig.~\ref{ER_dynamics}.)

To study the above results in greater detail, we allow $\eta$ to vary from $0$ to $1$ while keeping all other parameters fixed and focusing on the initial Poisson degree distribution. We compare simulation and AME results in Fig.~\ref{ER_dynamics}, illustrating one of the critical features of our model: the disease prevalence decreases with increasing $\eta$. That is, we observe that systems with a stronger preference for transitivity have smaller numbers of infected individuals, all else being equal. The thick lines in Fig.~\ref{ER_dynamics}(a) for the AME results accurately capture the final disease prevalence and roughly describe its temporal evolution   for $\eta<0.8$.  For $0.8 \leq \eta \leq 1$, AME underestimates the stationary disease prevalence $I$ but still captures qualitative features of the evolution. Remarking again that $\eta=1$ is a special case, we note that the system does not reach a well-defined stationary state in our simulations, with $I$ continuing to decrease.   In Fig.~\ref{eta_1}, we further explore the temporal evolution of $I$ when $\eta = 1$ while varying $\langle k \rangle$ from 2 to 10.  We identify a slow power-law decay in  $I$ for $\langle k\rangle=2$; using the curve fitting in MATLAB\textsuperscript{\textregistered}, we find that the disease prevalence $I$ for $\eta=1$ and $\langle k \rangle=2$ scales as $I=1.9{t}^{-0.173}$ (also shown in Fig.~\ref{eta_1}), with $I$ appearing to slowly approach a disease-free state.  As the mean degree increases, the observed decay with $t$ becomes even slower, to the point that we cannot claim anything about the functional form of the long-time decay from the Figure; nevertheless, we provide power-law fits to the late times in the Figure for comparison. Although the AME prediction for the temporal evolution for $\eta=1$ is qualitatively very different, it does appear to predict a correct $I_\infty=0$ disease-free stationary state [see Fig.~\ref{ER_dynamics}]. 

\begin{figure}
\centering
\includegraphics[width = 0.85\columnwidth]{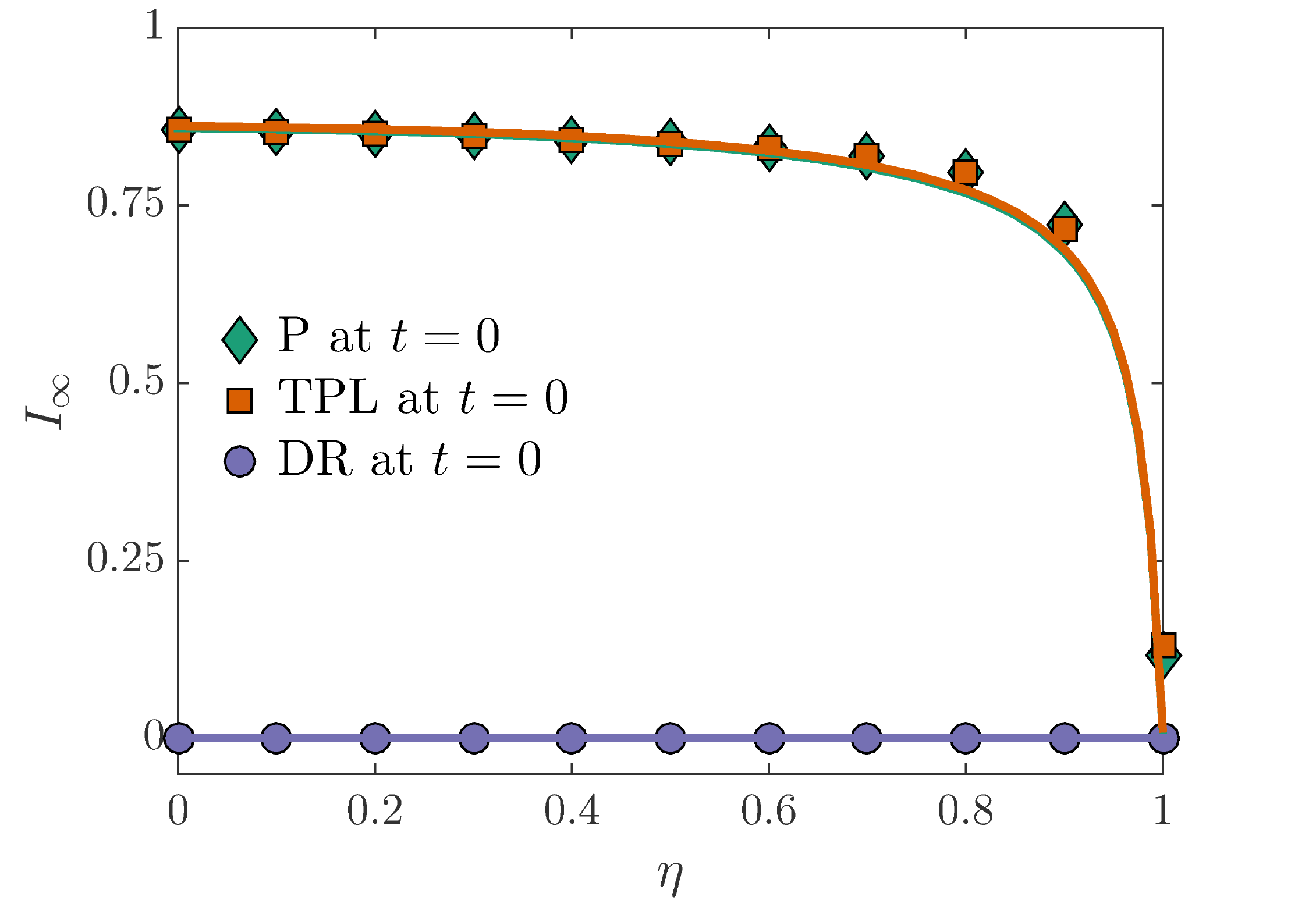}
\caption{(a) Disease prevalence in the stationary state $I_{\infty}$ versus $\eta$. The $I_{\infty}$'s were calculated at time $t = 10,000$. Initial degree distributions are Poisson (diamonds, $P$ at $t=0$, see Eq.~\ref{eq:pkp}), truncated power law (squares, $TPL$ at $t=0$, see Eq.~\ref{eq:tpl}) and degree regular (circles, $DR$ at $t=0$, see Eq.~\ref{eq:drd}). Markers correspond to means computed over 30 simulations and the lines are AME results. The other parameters are $\beta = 0.04$, $\gamma = 0.04$, $\alpha = 0.005$, $\epsilon = 0.1$ and  $\langle k \rangle = 2$.}
\label{disease_time_final}
\end{figure} 

\begin{figure}
\centering
\includegraphics[width = 0.5\textwidth]{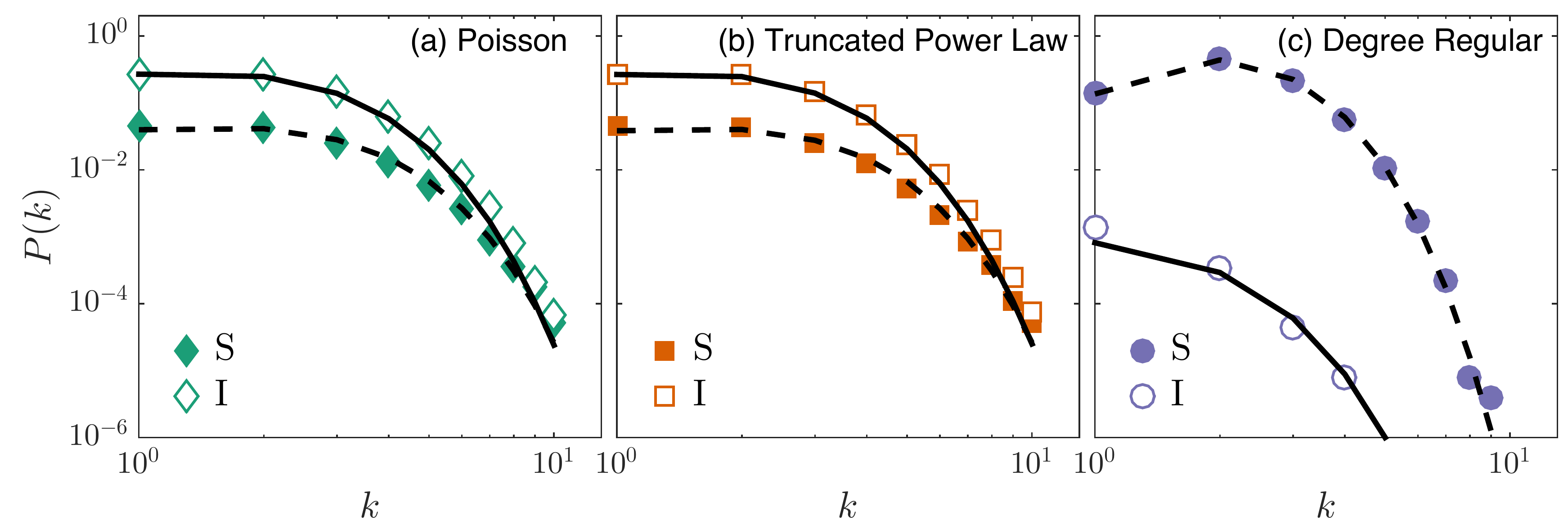}
\caption{Joint distributions of node states ($S$ and $I$) and degrees at stationarity. Initial degree distributions are (a) Poisson, (b) truncated power law and (c) degree regular. Markers correspond to means computed over 200 simulations. Dashed and solid lines are the AME predictions for the $S$ and $I$ states, respectively. Parameters used are $\beta = 0.04$, $\gamma = 0.04$, $\alpha = 0.005$, $\epsilon = 0.1$, $\eta = 0.4$ and $\langle k \rangle = 2$.}
\label{degree_distribution_1}
\end{figure} 

\begin{figure}
\centering
\includegraphics[width = \columnwidth]{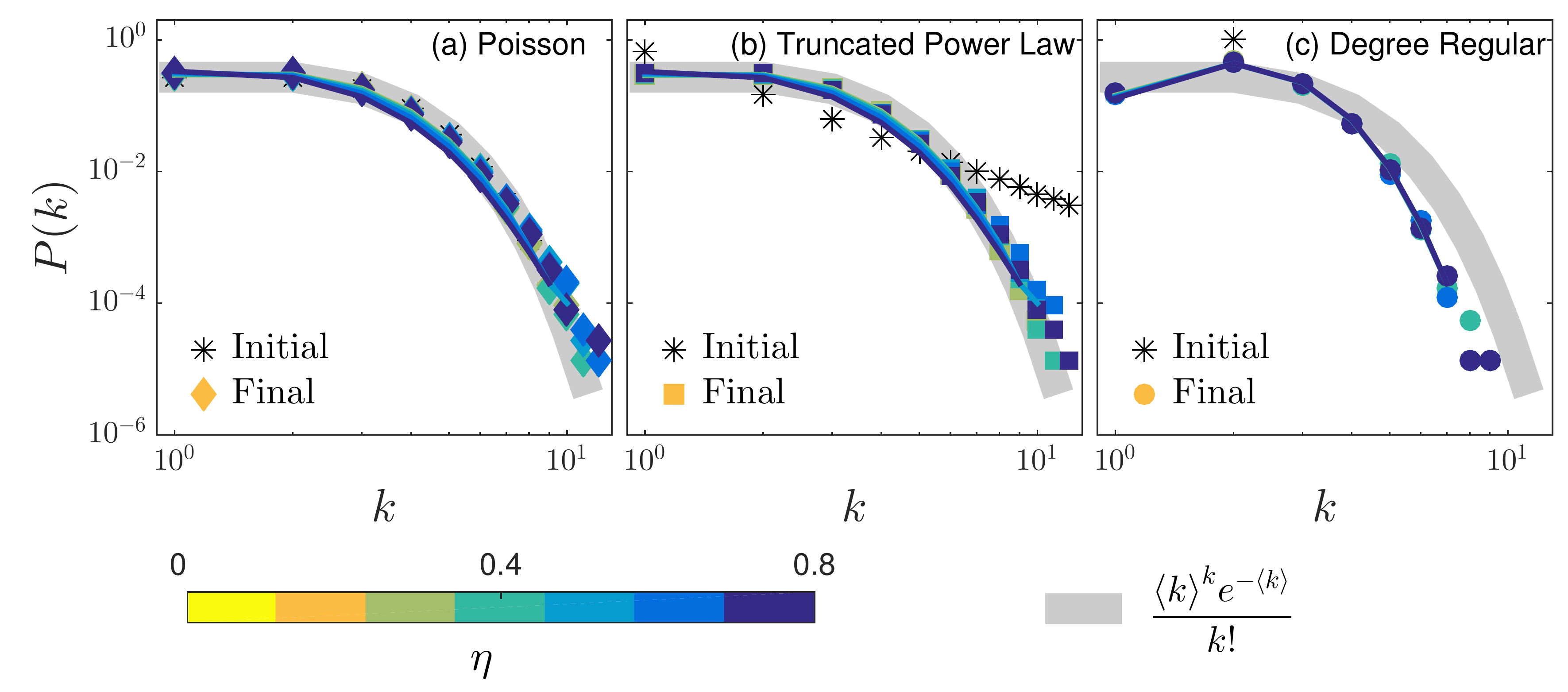}
\caption{Degree distributions in stationary states. Initial degree distributions are (a) Poisson, (b) truncated power law and (c) degree regular. Black asterisks in each subplot are the initial degree distributions. We plot results for $\eta = 0$, $0.2$, $0.4$, $0.6$, and $0.8$, as indicated by the color bar, though in many cases the results at one value of $\eta$ are obscured by those at other values. The other parameters are $\beta = 0.04$, $\gamma = 0.04$, $\alpha = 0.005$,  $\epsilon = 0.1$ and $\langle k \rangle = 2$. We plot results at $t= 5,000$, with all systems at these $\eta$ reaching their stationary states by this time. (Note that we did not plot $\eta = 1.0$ here.) Markers correspond to means computed over 200 simulations. Lines are the AME predictions for the stationary degree distributions. The thick gray line in the background is the reference Poisson distribution expected for a network with $\langle k \rangle = 2$. Observe that for both the initial Poisson and truncated power law the stationary state degree distribution falls onto this thick line, implying the final degree in both these cases is Poisson (at least approximately).}
\label{degree_distribution_2}
\end{figure}

\begin{figure*}%
  \centering
\includegraphics[width =1.75\columnwidth]{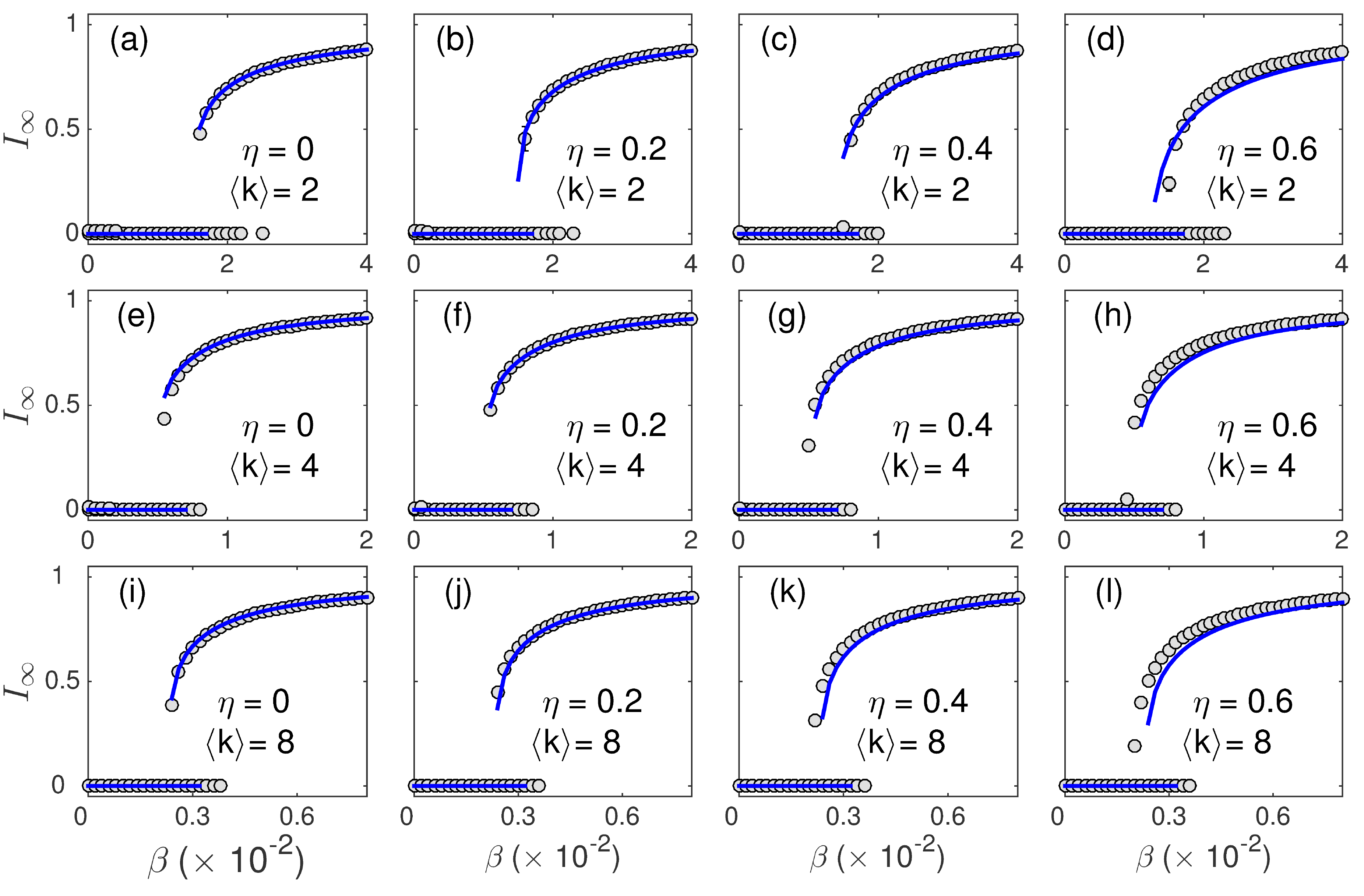}
   \caption{Bifurcation diagrams of stationary disease prevalence level $I_{\infty}$ versus infection rate $\beta$ on adaptive networks with an initial Poisson degree distribution. (1) $\langle k \rangle = 2$ for the first row of plots; (2) $\langle k \rangle = 4$ for the second row and (3) $\langle k \rangle = 8$ for the last row. In each row, we plot the case when $\eta = 0$, $0.2$, $0.4$, and $0.6$. Other parameters in the system are $\gamma = 0.02$ and $\alpha = 0.005$. Solid lines are the predictions of our semi-analytic method and the markers represent the average of 30 simulations. We ran these Monte Carlo simulations for each value $\epsilon = 0.001$, $0.01$, $0.05$, $0.99$ and $0.999$. For each run, the initial transient was discarded, and the prevalence at stationarity was averaged over at least $10,000$ time steps.}%
    \label{bifurcation_plot}%
\end{figure*}

To summarize, we plot simulation results (markers) and AME-based estimates (lines) of $I_{\infty}$ versus $\eta$ for different initial degree distributions in Fig.~\ref{disease_time_final}. The AME predictions and simulation results agree very well. The role of reinforced transitivity is clear: the final disease prevalence decreases with increasing transitivity reinforcement for networks starting with Poisson and power-law degree distributions. (The degree-regular cases are already disease-free at stationary for these parameters.)

In addition to predictions of disease prevalence, the AME approach contains and thus predicts the degree distributions of the coevolving networks. In Fig.~\ref{degree_distribution_1}, we plot the stationary joint distributions of node states and degrees for $\eta=0.4$ for the three different initial distributions; that is, the degree distributions of the susceptible $S$ and infected $I$ nodes are indicated separately with normalization corresponding to the respective fraction of nodes in each state. Markers and lines in Fig.~\ref{degree_distribution_1} represent the simulations and AME results, respectively, demonstrating that AME provides a good estimate in each case. Note that in  Fig.~\ref{degree_distribution_1}(c), the fraction of infected nodes is very close to but not exactly zero. This is because we stopped the numerical experiments when there are no $SI$ edge in the system; however, there are still a few $II$ edges and isolated infected nodes and as a result we still see infected nodes with different degrees in the stationary states.

To determine how $\eta$ impacts the degree distribution, we plot the initial and stationary degree distributions for different $\eta $'s in Fig.~\ref{degree_distribution_2}. As in previous figures, markers are simulations and lines are AME results, which appear to be in good agreement. Different colors represent different values of $\eta$, with the interesting result that $\eta$ appears to have only small effect on the total stationary degree distribution, even though we have seen before that $\eta$ does affect the disease prevalence. Indeed, for both the initial Poisson and truncated power law cases, the stationary degree distribution appears to be close to Poisson with mean degree $\langle k \rangle=2$ [see the thick gray line in  
 Fig.~\ref{degree_distribution_2}(a-b)]. This observation implies that the stationary states resulting from this process have both a Poisson degree distribution and nonzero transitivity. We recall that the model includes two kinds of rewiring: at random to any susceptible node and to a distance-two neighbor. Neither of these have direct preference to rewire to vertices with a particular degree, except that rewiring to a distance- two neighbor cannot rewire to a singleton. That is, both rewiring mechanisms are essentially similar to random rewiring in ignoring node degree, which leads to a Poisson degree distribution. Hence, we expect the initial degree distributions to converge to Poisson-like distributions in stationary.
 
 \begin{figure}
 \centering
    \includegraphics[width=\columnwidth]{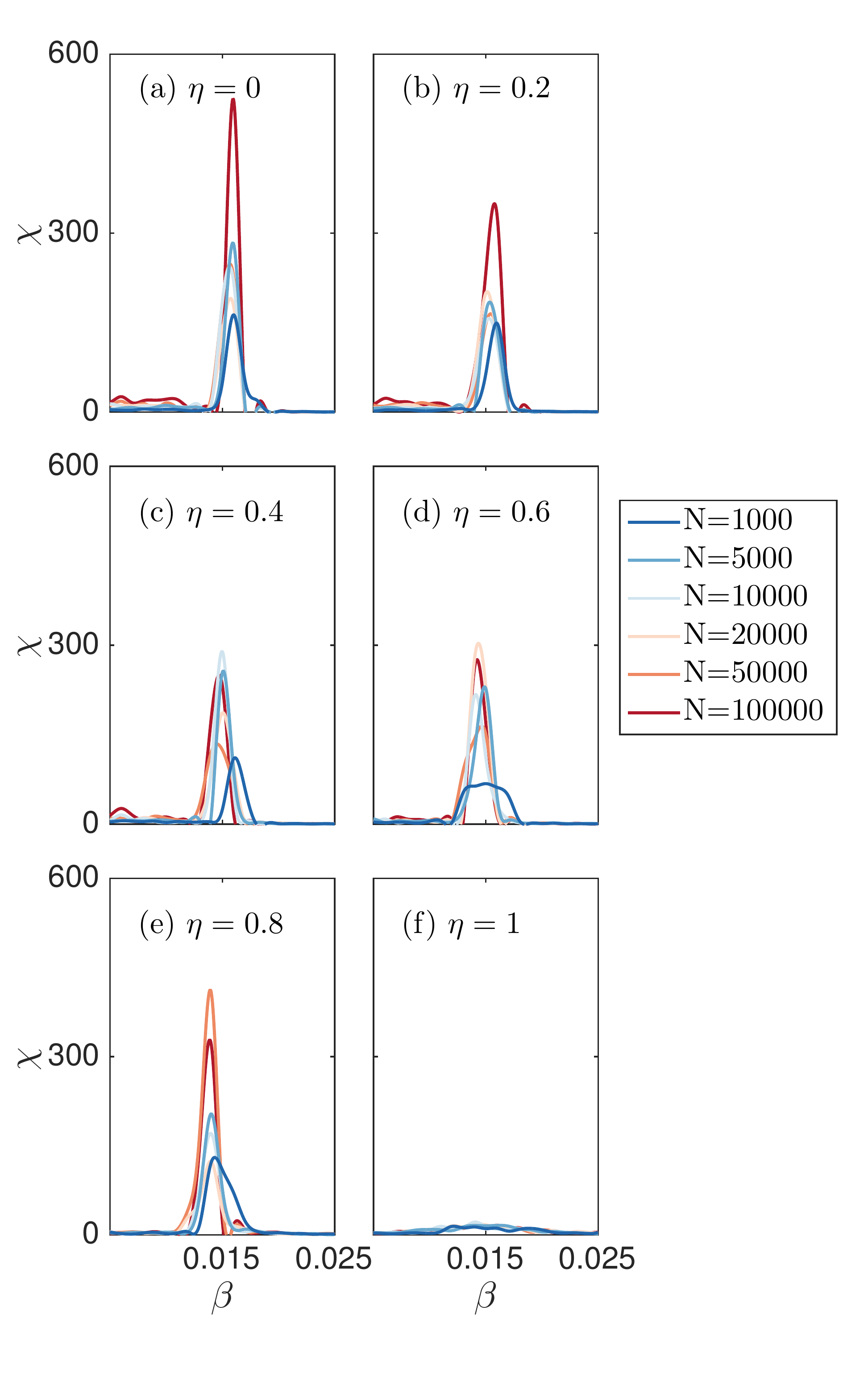}
    \caption{Susceptibility $\chi$ versus infected rate $\beta$ in adaptive networks with Poisson initial degree distributions with different network size $N$ at time $t = 10,000$ for various values of $\eta$: (a) $\eta = 0$, (b) $\eta = 0.2$, (c) $\eta = 0.4$, (d) $\eta = 0.6$, (e) $\eta = 0.8$, and (f) $\eta = 1$. Except for the case $\eta = 1$, all networks reached their stationary disease prevalence $I_{\infty}$. The other parameters of the system are $\gamma = 0.02$, $\alpha = 0.005$, $\epsilon = 0.1$, and $\langle k \rangle = 2$. Lines with different colors correspond to means computed over 30 Monte Carlo simulations of different network sizes.}
    \label{susceptibility}
\end{figure} 

\begin{figure}[!ht]
 \centering
    \includegraphics[width=\columnwidth]{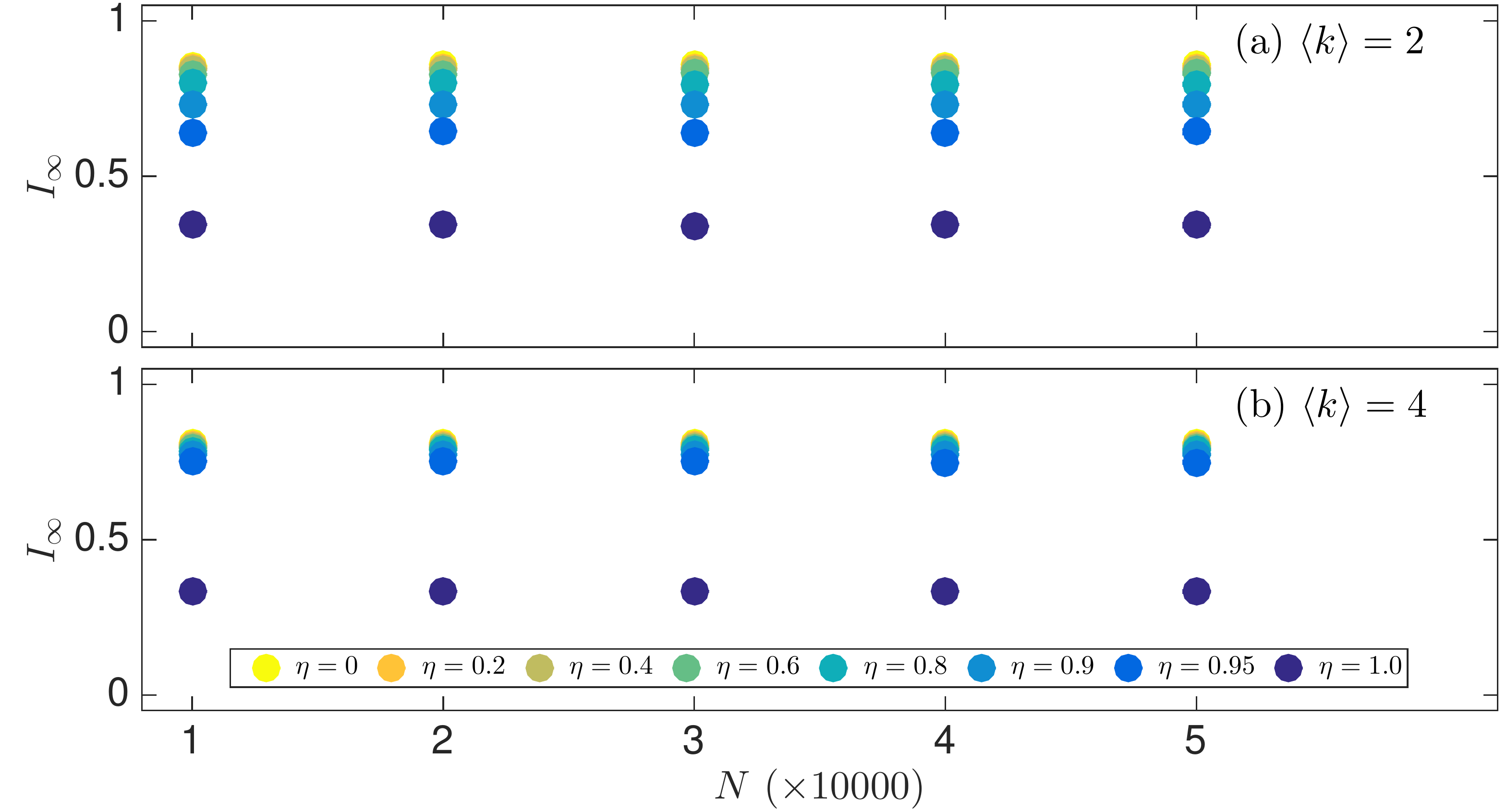}
    \caption{Stationary disease prevalence level $I_{\infty}$ (as measured at stopping time $t = 20,000$) versus network size on adaptive networks with an initial Poisson degree distribution with (a) $\langle k \rangle = 2$ and (b) $\langle k \rangle = 4$. In (a), the infected and rewiring rates chosen are: $\beta = 0.04$ and $\gamma = 0.04$. In (b), the infected and rewiring rates chosen are: $\beta = 0.01$ and $\gamma = 0.02$. Other parameters in the system are $\alpha = 0.005$ and $\epsilon = 0.1$. Dots (and error bars) represent the mean (and standard deviation) of outcomes from 30 Monte Carlo simulations. We vary $\eta$ from $0$ to $1$, recalling that $\eta = 1$ does not reach stationarity by $t = 20,000$. We note that $I_\infty$ is consistent across network sizes in the region tested in this paper.}
    \label{finite_size}
\end{figure}

\begin{figure}
    \centering
     \includegraphics[width=\columnwidth]{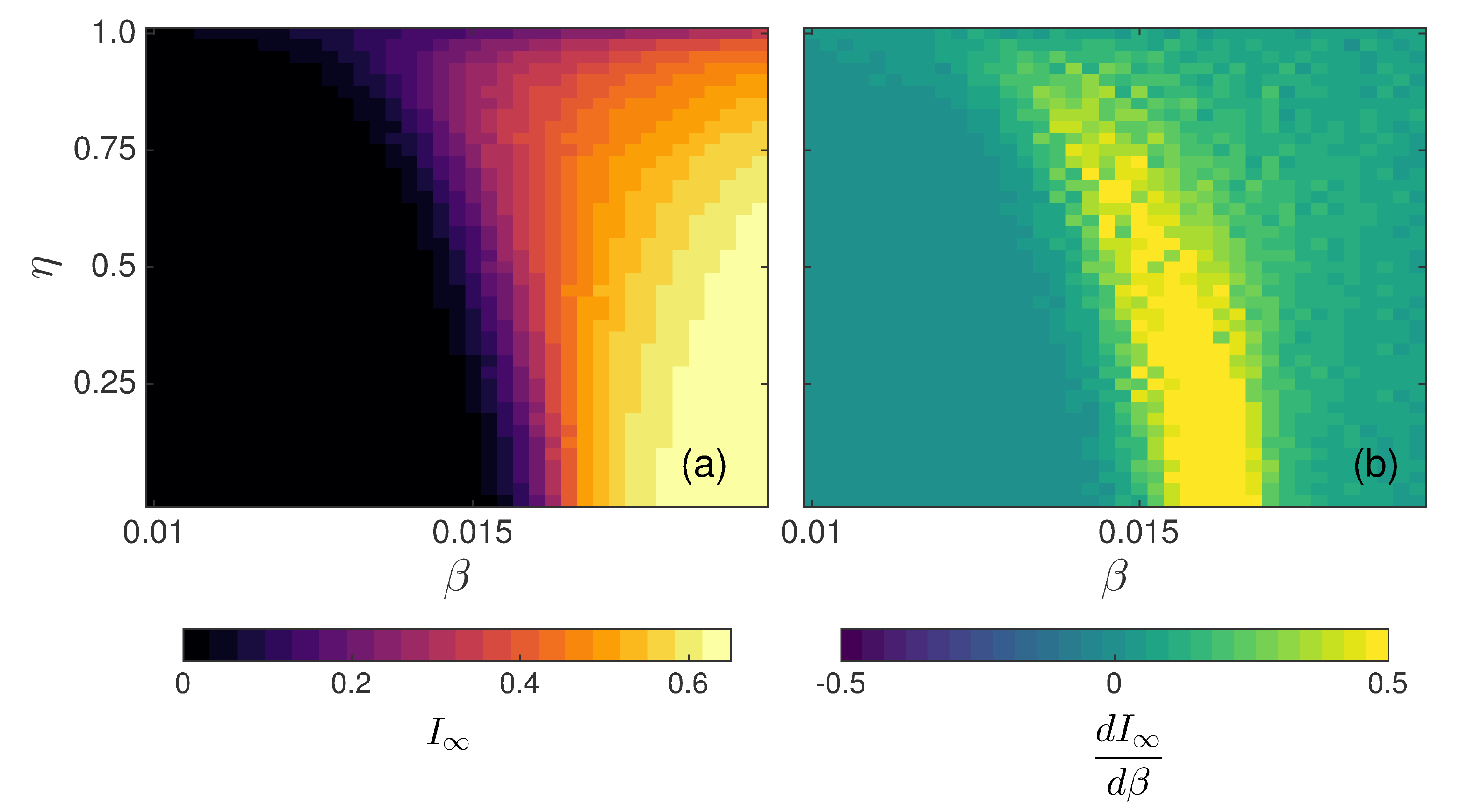}
    \caption{ (a) Phase diagram in parameter space $(\beta, \eta)$ for the observed disease prevalence $I$ at time $t = 10,000$ on the networks with Poisson initial degree distributions. (b) Derivative of disease prevalence with respect to $\beta$, i.e., $\frac{dI_\infty}{d\beta}$. The jump in $\frac{dI_\infty}{d\beta}$  with positive values implies a possible discontinuity of $\frac{dI_\infty}{d\beta}$.  Except for the case $\eta = 1$, all networks reached their stationary disease prevalence $I_{\infty}$.     
    We here fix the parameters of the system as $\gamma = 0.02$, $\alpha = 0.005$, $\epsilon = 0.1$, and $\langle k \rangle = 2$. Each pixel corresponds to the mean computed over 30 Monte Carlo simulations.}
    \label{phase_diagram}
\end{figure}

Similar to Figure 5 in Ref.~\cite{marceau2010adaptive}, we plot (Fig.~\ref{bifurcation_plot}) bifurcation diagrams of stationary disease prevalence level $I_{\infty}$ versus infection rate $\beta$ with an initial Poisson degree distribution and with different mean degree $\langle k \rangle = 2$, $4$, and $8$. The other parameters here include $\gamma = 0.02$ and $\alpha = 0.005$. We obtained the average stationary disease prevalence level $I_{\infty}$ at each value of $\beta$ from 30 Monte Carlo simulations for each initial infected proportion $\epsilon = 0.001$, $0.01$, $0.05$, $0.99$, and $0.999$. We call attention to the range of the infection rate $\beta$ plotted in Fig.\ref{bifurcation_plot}, ranging from $0$ to $0.04$ in panels (a)--(d) for $\langle k\rangle=2$, from $0$ to $0.02$ in panels (e)--(h) for $\langle k\rangle=4$, and from $0$ to $0.008$ in panels (i)--(l) for $\langle k\rangle=8$. As evident in the figure, our AME calculation reasonably predicts the stationary disease prevalence level $I_{\infty}$, including its phase transition and the bistability near that transition.
 
We observed that varying $\eta$  (the probability of closing triangles during rewiring) can lead to the transition of the system from endemic to disease-free state with a very slow time scale when $\eta = 1$ (see Fig.~\ref{eta_1}). To further investigate whether the nature of these transitions might be a finite size effect, in Fig.~\ref{susceptibility} we plot the susceptibility $\chi = N (\langle I_{\infty}^2 \rangle - \langle I_{\infty} \rangle^{2})/ \langle I_{\infty} \rangle$, as  defined in \cite{ferreira2012epidemic}, for different  network size $N$. With the peak of the susceptibility identifying the transition, the results in the figure show that for $\eta<1$ the network size $N$ does not greatly change the transition location in $\beta$, except for in the smallest networks. However, we again see a qualitative difference for $\eta = 1$, making a definitive conclusion in this case more difficult.

In Fig.~\ref{finite_size} we consider a complementary test of possible finite-size effects: keeping the average degree fixed (with initial Poisson degree distributions), we vary the network sizes from $N = 10,000$ to $N = 50,000$ while considering different values of $\eta$ from 0 to 1. Fixing the average degree, the final disease prevalence level appears to be almost constant across network sizes. That said, we note that the $\beta$ values considered in Fig.~\ref{finite_size} are clearly above the transition between disease-free and endemic cases, though the $\beta$ used with $\langle k\rangle=4$ in Fig.~\ref{finite_size}(b) is much closer to the transition than that in Fig.~\ref{finite_size}(a). Combined with Fig. 14, where the susceptibility $\chi$ and phase transition seem to not depend on the network size, these findings suggest that, in general, the stationary disease prevalence level $I_{\infty}$ does not change appreciably with the size of the system here. While we acknowledge that the precise details in and immediately around the transition in $\beta$ might be expected to be more sensitive to system size, such an exploration would involve an even more intensive numerical investigation simultaneously considering variations in $N$, $\beta$ and $\eta$. Meanwhile, the above results together appear to support the hypothesis that the qualitative nature of the transitions observed herein at various parameter settings are not likely to be consequences of finite system sizes.

Furthermore, to demonstrate the influence of $\eta$ on the transitions, in Fig.~\ref{phase_diagram} (a) we plot a phase diagram for the observed stationary disease prevalence $I_{\infty}$ in the parameter space $(\beta, \eta)$. We observe that as $\eta$ increases, the critical value of $\beta$ where the system changes from disease-free to endemic decreases.  To supplement the phase diagram,  in Fig.~\ref{phase_diagram} (b) we plot the numerically-computed derivative of disease prevalence along the $\beta$ direction, ${dI_\infty}/{d\beta}$, observing only a narrow range with large positive value near the transition points apparent in (a). Also, the jump in  $\frac{dI_\infty}{d\beta}$ occurs at a lower $\beta$ as $\eta$ increases, again showing the dependence of the transitions on $\eta$. Lastly, when $\eta = 1$, we see no jumps of $I_{\infty}$ when $\beta$ varies, further suggesting that the system behavior at $\eta = 1$ might be qualitatively different. The observed transitions might also depend on other parameters in the system. In Appendix A.2, we study the influence of $\epsilon$ (initial fraction of infected nodes) on the transitions, and the effect of $\eta$ remains.

\begin{figure*}
\centering
\includegraphics[width =1.75\columnwidth]{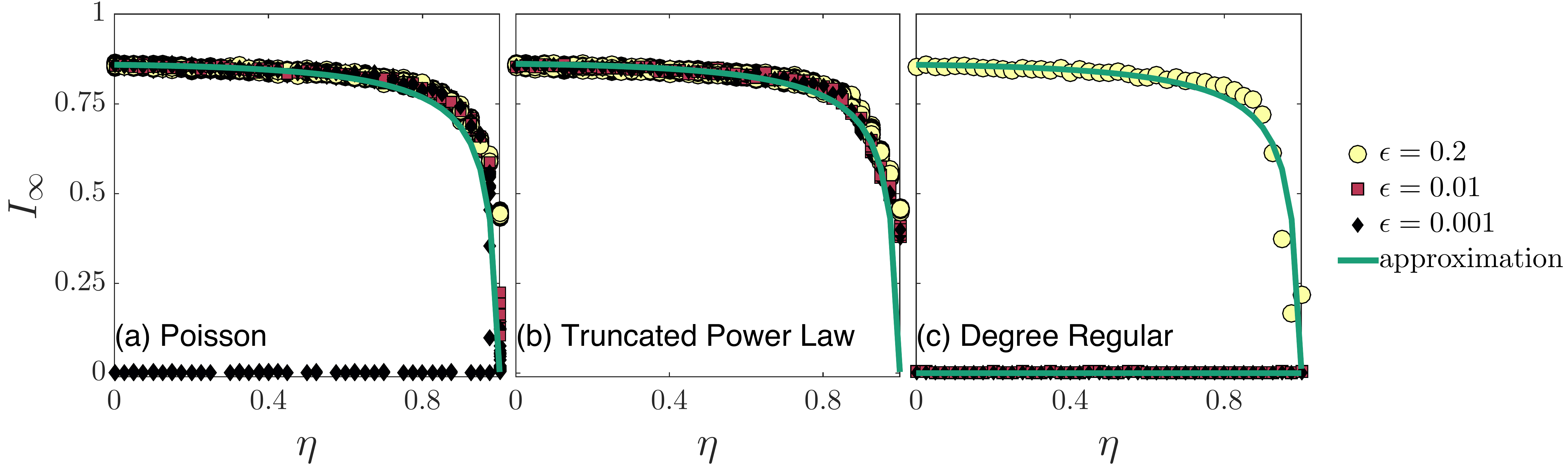}
\caption{Bifurcation diagrams of the disease prevalence $I$ versus $\eta$ on networks with initial degree distributions that are (a) Poisson, (b) truncated power law, and (c) degree regular. The parameters of the system are $\beta = 0.04$, $\gamma = 0.04$, $\alpha = 0.005$, and $\langle k \rangle = 2$. Lines are the AME predictions and dots are the means from 30 simulations. We run simulations for each value $\epsilon = 0.001, 0.01, 0.2, 0.4, 0.6, 0.8, 0.99$, and, $0.999$ and plot them with different colors. We set $t= 10,000$, with all parameters with $\eta$ not close to 1 reaching their stationary states by this time.}
\label{bifurcation_2}
\end{figure*}

Finally, we study the effect of $\eta$ on the observed bifurcations while varying the fraction of initial infections, $\epsilon$. Instead of using a fixed value of $\epsilon$ (we recall $\epsilon=0.1$ in the most recent sets of figures above), we consider $\epsilon$ to be $0.001$, $0.01$, $0.2$, $0.4$, $0.6$, $0.8$, $0.99$, and $0.999$. We run simulations to $t = 10,000$ and once again observe stationary states for $\eta<1$. In Fig.~\ref{bifurcation_2}, we show that AME predicts the disease prevalence curves for the three different initial degree distributions. Fig.~\ref{bifurcation_2} also illustrates the effect of $\epsilon$ on the system. For example, in the degree-regular case, the infection becomes endemic only for $\epsilon > 0.1$, whereas in the truncated power law case every $\epsilon$ leads to an endemic state. In contrast, the Poisson case is more involved: the system appears to be \emph{bistable}, with larger $\epsilon$ yielding an endemic infection and smaller $\epsilon$ typically reaching a disease-free state. Even so, we note that the $\epsilon=0.001$ points in the figure include points on both the disease-free and endemic branches for the Poisson case.

We further investigate this behavior with calculations at $\epsilon = 0.00004$, so on average, we expect on average to initially see only one infected node, and with probability $(1-0.00004)^{25000} \doteq 37$\% the networks start with no infected nodes. We performed 30 Monte Carlo simulations for this $\epsilon$, finding that 80\% of the networks lead to a disease-free state for Poisson initial condition; on the other hand, 60\% of the networks lead to a disease-free state for truncated power law initial condition. We observe that when $\epsilon$ is small enough, the system can converge to either an endemic or disease-free state. AME is unable to predict this bistability from a fixed $\epsilon$. However, in the cases where the system exhibits only one type of stationary state, then our AME system gives a good prediction of the stationary disease prevalence $I_{\infty}$. In the degree-regular case [see Fig.~\ref{bifurcation_2}(c)], there is an additional horizontal line indicating that AME can capture the state when $I_\infty = 0$ for $\epsilon \leq 0.1$.  In Fig.~\ref{bifurcation_2}, there is greater variability near $\eta=1$ because the system does not converge to a stationary state for the simulated times. Fig.~\ref{bifurcation_2} also shows that for large enough $\epsilon$, $I_\infty$ becomes independent of the initial degree distribution, with the results falling on the same universal curve.

\section{Discussion}

We have introduced a new model variant for the spread of disease on a coevolving network with reinforcement of transitivity---a quintessential property of social networks. Transitivity in the model is controlled by a combination of the rewiring rate $\gamma$ and an additional parameter $\eta$ for the probability of closing triangles. This model provides an opportunity to study the role of transitivity in altering the dynamics of disease spread.  

We explored the parameter space of $\gamma$ and $\eta$ with three different initial degree distributions. We have identified that higher values of $\eta$  lead to lower disease prevalence. In other words, increased reinforcement of transitivity can decrease the disease prevalence. This finding suggests a possible general mechanism for controlling SIS-type epidemics, intuitively encouraging healthy individuals to focus their contacts to healthy individuals within a close social circle, and this will typically do better than building contacts to random people, in the sense of leading to lower endemic rates. The viability of this general intuition beyond the specific setting of direct closure of triangles may be important for future study.

We carried out a bifurcation analysis to understand the properties of the systems at equilibrium. Fig.~\ref{bifurcation_1} and Appendix A.2 implies that the phase transition from disease-free to epidemic states as $\beta$ varies might be discontinuous when $\eta<1$ and continuous when $\eta = 1$. Validating the nature of this change is a promising potential direction for further studies.
 
We extended the AME method to include the effect of transitivity reinforcement. We showed that for $\eta < 0.8$, our AME system predicts the disease prevalence in the stationary state. Furthermore, we illustrated the accuracies of the stationary degree distributions predicted by AME. This success of AME further supports its use to study a variety of binary state dynamics on coevolving networks, accurately predicting properties of large networks at a manageable computational cost. We remark that the high AME accuracy observed here for non-zero $\eta$ may be surprising, in that a key assumption of the AME method is that the networks are locally tree-like, but this assumption will seemingly become less valid as local clustering increases with increasing $\eta$. Nevertheless, our AME method appears able to provide good predictions; the possible reasons for such accuracy remain a potential area for future exploration. (See, e.g., Ref.~\cite{melnik2011unreasonable} for other settings where theoretical models remain good predictions in the presence of significant local clustering.)

The behavior of this model at $\eta=1$ (i.e., only rewiring to close triangles) is different from the $\eta<1$ cases. When $\eta=1$, the system evolves very slowly and does not converge to a stationary state in our simulated times, with the disease prevalence appearing to decay as a small inverse power of $t$. We surmise that the system is slowly approaching a disease-free state. In this case, AME is not able to quantitatively describe the temporal evolution of the system; however, it appears to still correctly predict the final disease-free state. 

\section*{Acknowledgements}
Research reported in this publication was supported by the Eunice Kennedy Shriver National Institute of Child Health \& Human Development of the National Institutes of Health under Award Number R01HD075712. Additional support was provided from the James S. McDonnell Foundation 21st Century Science Initiative - Complex Systems Scholar Award grant \#220020315. The content is solely the responsibility of the authors and does not necessarily represent the official views of the institutions supporting this work.

\nocite{*}
\bibliographystyle{unsrt}

\begin{thebibliography}{31}

\bibitem{keeling2005networks}
Matt~J. Keeling and Ken~T. D. Eames.
\newblock Networks and epidemic models.
\newblock {\em Journal of the Royal Society Interface}, 2(4):295--307, 2005.

\bibitem{newman2002spread}
Mark~E. J. Newman.
\newblock Spread of epidemic disease on networks.
\newblock {\em Physical review E}, 66(1):016128, 2002.

\bibitem{gross2008adaptive}
Thilo Gross and Bernd Blasius.
\newblock Adaptive coevolutionary networks: a review.
\newblock {\em Journal of The Royal Society Interface}, 5(20):259--271, 2008.

\bibitem{sayama2013modeling}
Hiroki Sayama, Irene Pestov, Jeffrey Schmidt, Benjamin~James Bush, Chun Wong,
  Junichi Yamanoi, and Thilo Gross.
\newblock Modeling complex systems with adaptive networks.
\newblock {\em Computers \& Mathematics with Applications}, 65(10):1645--1664,
  2013.
  
\bibitem{vazquez2007impact}
Alexei Vazquez, Bal\'azs R\'acz, Andr\'as Luk\'acs, and Albert-L\'aszl\'o Barab\'asi.
\newblock Impact of non-Poissonian activity patterns on spreading processes.
\newblock {\em Physical Review L }, 98(15): 158702, 2007.

\bibitem{karsai2011small}
Small but slow world: How network topology and burstiness slow down spreading.
\newblock M. Karsai, M. Kivel\"a, R. K. Pan, K. Kaski, J. Kert\'esz, A.-L. Barab\'asi, and J. Saram\"aki.
\newblock {\em Physical Review E }, 83(2): 025102, 2011.  

\bibitem{vestergaard2014memory}
Christian L. Vestergaard, Mathieu G{\'e}nois, and Alain Barrat.
\newblock How memory generates heterogeneous dynamics in temporal networks.
\newblock {\em Physical Review E }, 90(4):042805, 2014.
  
\bibitem{holme2014birth}
Birth and death of links control disease spreading in empirical contact networks.
\newblock Petter Holme and Fredrik Liljeros.
\newblock {\em Scientific Reports}, 4: 4999, 2014.

\bibitem{Regenbogen6400}
Behavioral and neural correlates to multisensory detection of sick humans
\newblock   Christina Regenbogen,  John Axelsson,  Julie  Lasselin,  Danja K Porada,  Tina Sundelin,  Moa G. Peter,  Mats  Lekander,  Johan N Lundstr{\"o}m, and Mats J. Olsson. 
\newblock{ \em PNAS},  114(24): 6400, 2017

\bibitem{read2008dynamic}
Jonathan~M. Read, Ken~T. D. Eames, and W.~John Edmunds.
\newblock Dynamic social networks and the implications for the spread of
  infectious disease.
\newblock {\em Journal of The Royal Society Interface}, 5(26):1001--1007, 2008.

\bibitem{hill2010infectious}
Alison~L. Hill, David~G. Rand, Martin~A. Nowak, and Nicholas~A. Christakis.
\newblock Infectious disease modeling of social contagion in networks.
\newblock {\em PLoS Comput Biol}, 6(11):e1000968, 2010.

\bibitem{shaw2008fluctuating}
Leah~B. Shaw and Ira~B. Schwartz.
\newblock Fluctuating epidemics on adaptive networks.
\newblock {\em Physical Review E}, 77(6):066101, 2008.

\bibitem{zanette2008infection}
Dami{\'a}n~H. Zanette and Sebasti{\'a}n Risau-Gusm{\'a}n.
\newblock Infection spreading in a population with evolving contacts.
\newblock {\em Journal of biological physics}, 34(1-2):135--148, 2008.

\bibitem{funk2010modelling}
Sebastian Funk, Marcel Salath{\'e}, and Vincent~A. A. Jansen.
\newblock Modelling the influence of human behaviour on the spread of
  infectious diseases: a review.
\newblock {\em Journal of the Royal Society Interface}, 7(50):1247--1256, 2010.

\bibitem{moore2000epidemics}
Cristopher Moore and Mark~E. J. Newman.
\newblock Epidemics and percolation in small-world networks.
\newblock {\em Physical Review E}, 61(5):5678, 2000.

\bibitem{miller2009percolation}
Joel~C. Miller.
\newblock Percolation and epidemics in random clustered networks.
\newblock {\em Physical Review E}, 80(2):020901, 2009.

\bibitem{kuperman2001small}
Marcelo Kuperman and Guillermo Abramson.
\newblock Small world effect in an epidemiological model.
\newblock {\em Physical Review Letters}, 86(13):2909, 2001.

\bibitem{hufnagel2004forecast}
Lars Hufnagel, Dirk Brockmann, and Theo Geisel.
\newblock Forecast and control of epidemics in a globalized world.
\newblock {\em Proceedings of the National Academy of Sciences of the United
  States of America}, 101(42):15124--15129, 2004.

\bibitem{newman2009random}
Mark~E. J. Newman.
\newblock Random graphs with clustering.
\newblock {\em Physical review letters}, 103(5):058701, 2009.

\bibitem{gleeson2009bond}
James~P. Gleeson.
\newblock Bond percolation on a class of clustered random networks.
\newblock {\em Physical Review E}, 80(3):036107, 2009.

\bibitem{saeedian2017epidemic}
M. Saeedian, N. Azimi-Tafreshi, G. R. Jafari, and J. Kertesz.
\newblock Epidemic spreading on evolving signed networks.
\newblock {\em Physical Review E}, 95(2):022314, 2017.

\bibitem{ebel2002dynamics}
Holger Ebel, J{\"o}rn Davidsen, and Stefan Bornholdt.
\newblock Dynamics of social networks.
\newblock {\em Complexity}, 8(2):24--27, 2002.

\bibitem{holme2006nonequilibrium}
Petter Holme and Mark~E. J. Newman.
\newblock Nonequilibrium phase transition in the coevolution of networks and
  opinions.
\newblock {\em Physical Review E}, 74(5):056108, 2006.

\bibitem{scarpino2016effect}
Samuel~V. Scarpino, Antoine Allard, and Laurent H{\'e}bert-Dufresne.
\newblock The effect of a prudent adaptive behaviour on disease transmission.
\newblock {\em Nature Physics}, 2016.

\bibitem{malik2013role}
Nishant Malik and Peter~J. Mucha.
\newblock Role of social environment and social clustering in spread of
  opinions in coevolving networks.
\newblock {\em Chaos: An Interdisciplinary Journal of Nonlinear Science},
  23(4):043123, 2013.

\bibitem{malik2016transitivity}
Nishant Malik, Feng Shi, Hsuan-Wei Lee, and Peter~J. Mucha.
\newblock Transitivity reinforcement in the coevolving voter model.
\newblock {\em Chaos: An Interdisciplinary Journal of Nonlinear Science},
  26(12):123112, 2016.

\bibitem{thurner2016multiplex}
Peter Klimek, Marina Diakonova, V{\'\i}ctor~M. Egu{\'\i}luz, Maxi~San Miguel, and Stefan Thurner.
\newblock Dynamical origins of the community structure of an online multi-layer
  society.
\newblock {\em New J. Phys.}, 18:083045, 2016.

\bibitem{gleeson2011high}
James~P. Gleeson.
\newblock High-accuracy approximation of binary-state dynamics on networks.
\newblock {\em Physical Review Letters}, 107(6):068701, 2011.

\bibitem{gleeson2013binary}
James~P. Gleeson.
\newblock Binary-state dynamics on complex networks: Pair approximation and beyond.
\newblock {\em Physical Review X}, 3(2):021004, 2013.

\bibitem{brauer2001mathematical}
Fred Brauer and Carlos Castillo-Chavez.
\newblock {\em Mathematical models in population biology and epidemiology},
  volume~1.
\newblock Springer, 2001.

\bibitem{gross2006epidemic}
Thilo Gross, Carlos J.~Dommar D'Lima, and Bernd Blasius.
\newblock Epidemic dynamics on an adaptive network.
\newblock {\em Physical review letters}, 96(20):208701, 2006.

\bibitem{zschaler2012largenet2}
Gerd Zschaler and Thilo Gross.
\newblock Largenet2: an object-oriented programming library for simulating large adaptive networks.
\newblock {\em Bioinformatics}, bts663, 2012.

\bibitem{gillespie1976general}
Daniel~T. Gillespie.
\newblock A general method for numerically simulating the stochastic time evolution of coupled chemical reactions.
\newblock {\em Journal of Computational Physics}, 22(4):403--434, 1976.

\bibitem{marceau2010adaptive}
Vincent Marceau, Pierre-Andr{\'e} No{\"e}l, Laurent H{\'e}bert-Dufresne,
  Antoine Allard, and Louis~J Dub{\'e}.
\newblock Adaptive networks: Coevolution of disease and topology.
\newblock {\em Physical Review E}, 82(3):036116, 2010.

\bibitem{watts1998collective}
Duncan J. Watts and Steven H. Strogatz.
\newblock Collective dynamics of ``small-world'' networks.
\newblock {\em Nature}, 393.6684 (1998): 440. 

\bibitem{wasserman1994social}
Wasserman Stanley, and Katherine Faust.
\newblock Social network analysis: Methods and applications.
\newblock {\em  Cambridge university press}, 1994, Vol. 8.

\bibitem{durrett2012graph}
Richard Durrett, James~P. Gleeson, Alun~L. Lloyd, Peter~J. Mucha, Feng Shi, David Sivakoff, Joshua~E. S. Socolar, and Chris Varghese.
\newblock Graph fission in an evolving voter model.
\newblock {\em Proceedings of the National Academy of Sciences},
  109(10):3682--3687, 2012.

\bibitem{zhou2013link}
Jie Zhou, Gaoxi Xiao, and Guanrong Chen.
\newblock Link-based formalism for time evolution of adaptive networks.
\newblock {\em Physical Review E}, 88(3):032808, 2013.

\bibitem{lee2017evol}
Hsuan-Wei Lee, Nishant Malik, and Peter J Mucha
\newblock Evolutionary prisoner's dilemma games coevolving on adaptive networks.
\newblock {\em Journal of Complex Networks}, 6(1): 1-23, 2017. 

\bibitem{ferreira2012epidemic}
Silvio C. Ferreira, Claudio Castellano, and Romualdo Pastor-Satorras.
\newblock Epidemic thresholds of the susceptible-infected-susceptible model on networks: A comparison of numerical and theoretical results.
\newblock {\em Physical Review E} 86(4): 041125, 2012.



\bibitem{melnik2011unreasonable}
Sergey Melnik, Adam Hackett, Mason A. Porter, Peter J. Mucha, and James P. Gleeson.
\newblock The unreasonable effectiveness of tree-based theory for networks with clustering.
\newblock {\em Physical Review E}, 83(3):036112, 2011. 








\end{thebibliography}

\appendix
\section{Appendices}

\begin{figure*}
    \centering
    \includegraphics[width = \columnwidth]{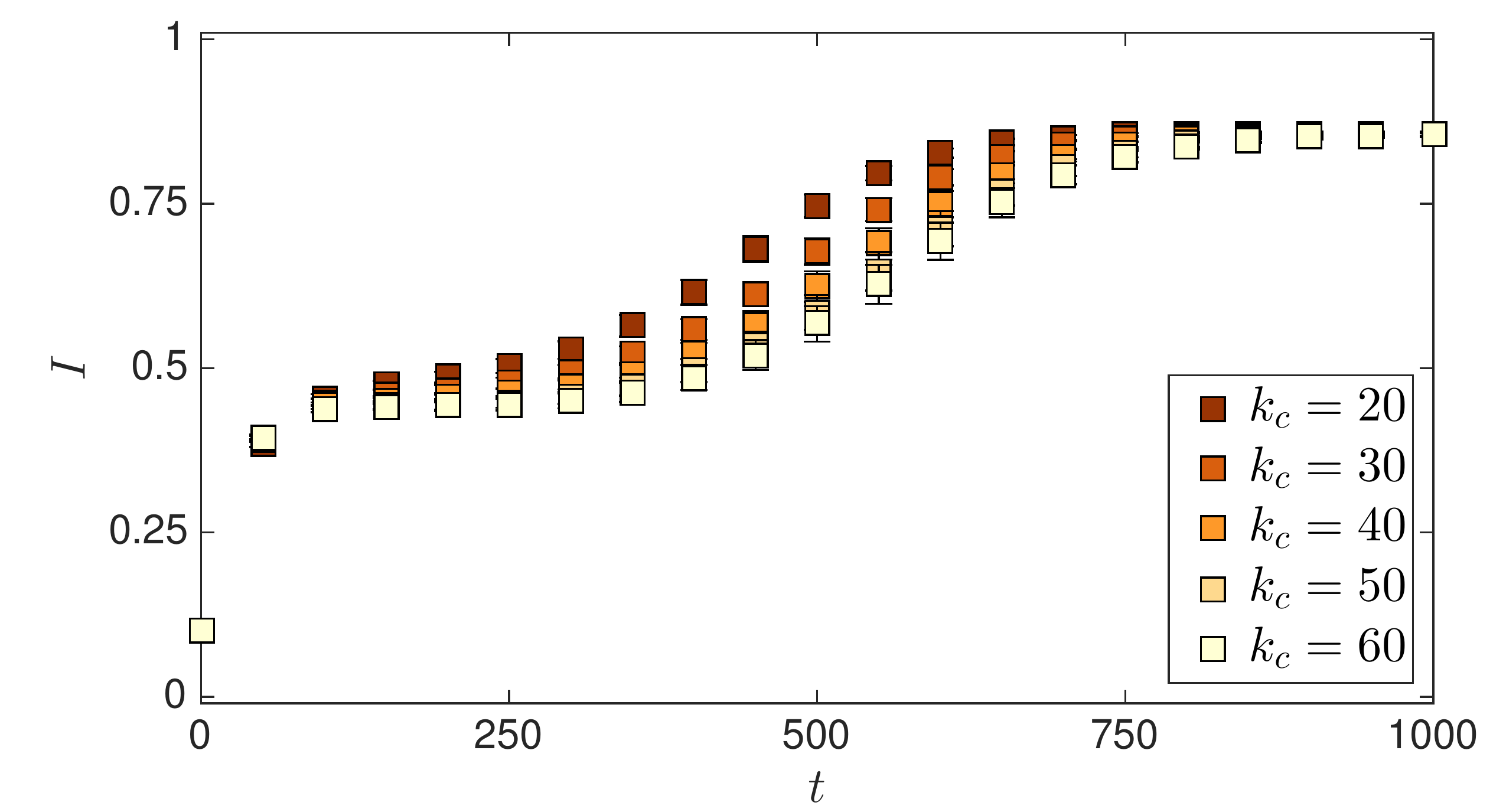} 
   \caption{Disease prevalence $I$ at time $t$ with different cut-off degrees $k_c$ in networks with power-law degree distribution at the start of the simulation  (see Eq.~\ref{eq:tpl}).   Markers correspond to means with standard deviations computed over 30 simulations. The parameter used are $\gamma = 0.04$, $\beta = 0.04$, $\alpha = 0.005$, $\eta = 0.2$ and $\epsilon = 0.1$. Different cut-off degrees in TPL have different levels of disease prevalence temporarily, however, as $t$ increases, they lead to the same level of disease prevalence. }%
    \label{truncation1}%
\end{figure*}

\subsection{Different cut-off degrees of truncated power law distribution}

Truncated power laws depend on an additional parameter, known as the cut-off degree, $k_c$ (see Eq.~\ref{eq:tpl} above for details).   Here, we examine the effects of $k_c$ on the variable $I$ (disease prevalence level ) at time $t$.  Specifically, we keep the mean degree $\langle k \rangle = 2$, and vary the cut-off degree $k_c$ from $20$ to $60$. We found that different cut-off degree of truncated power law distribution yield different level of disease prevalence level temporarily, but they have the same level of disease prevalence and degree distribution at stationarity.

\subsection{Bifurcation diagrams with different $\epsilon$}
\begin{figure*}
\centering
\includegraphics[width = 2\columnwidth]{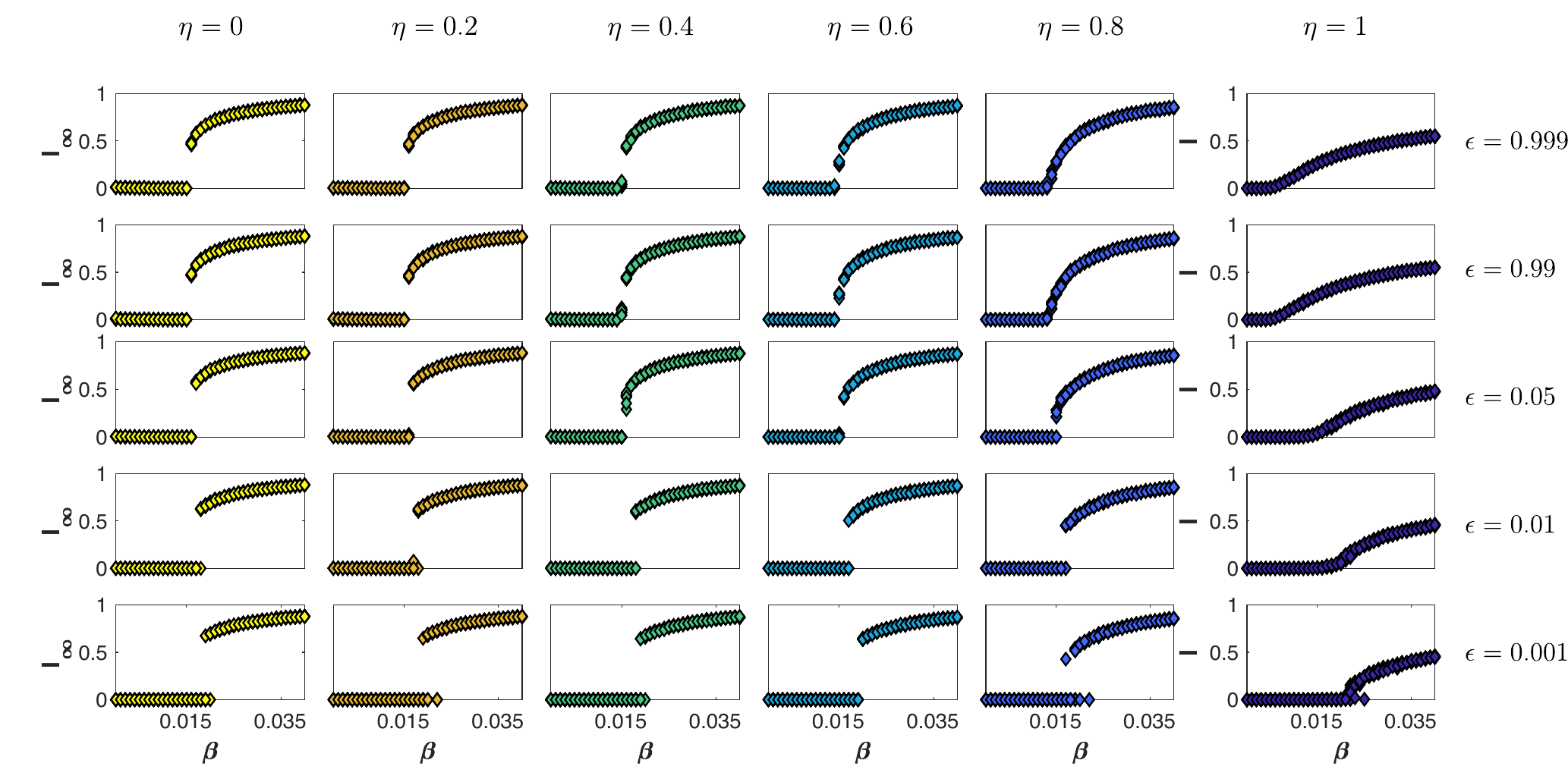}
    \caption{Bifurcation diagrams of the disease prevalence in the stationary state  $I_{\infty}$ versus $\beta$ at various values of $\eta$ and $\epsilon$. Each column of plots corresponds to a particular value of $\eta$ and each row of plots corresponds to a particular value of $\epsilon$. As in Fig.~\ref{bifurcation_1}, every point represents a mean of 30 realizations of the system at $t= 10,000$. The last column belongs to the $\eta=1$ case, noting that the $y$-axis is labeled $I$ (cf.\ $I_{\infty}$) because the simulations have not reached stationarity. The networks have a Poisson degree distribution initially. Other parameters of the system are $\gamma = 0.02$, and $\alpha = 0.005$.}
    \label{different_eps}
\end{figure*}

Here we revisit the bifurcation diagram in Fig.~\ref{bifurcation_1} and let the initial infected ratio $\epsilon$ also vary from 0.001 to 0.999. In Fig.~\ref{different_eps}, we can see that the critical region of transition also depends on $\epsilon$.

\subsection{AME and the transitory dynamics}

\begin{figure*}
 \vspace{0.1cm}
 \centering
    \includegraphics[width=\columnwidth ]{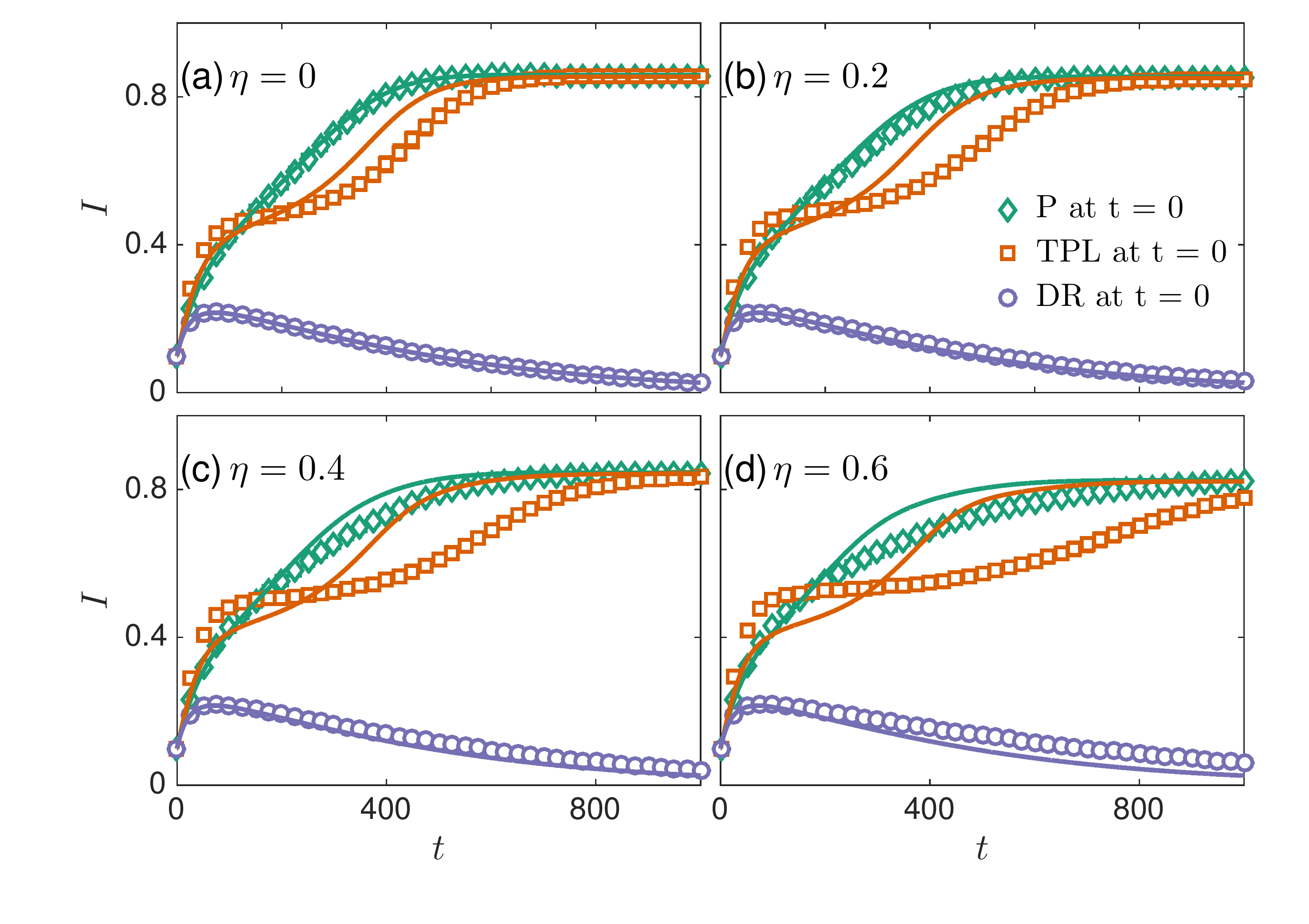}
    \caption{Early evolution of the disease prevalence $I$ in adaptive networks with Poisson ($P$), truncated power law (TPL), and degree regular (DR) initial degree distributions. The other parameters of the system are $\beta = 0.04$, $\gamma = 0.04$, $\alpha = 0.005$, $\epsilon = 0.1$, and $\langle k \rangle = 2$. Different values of $\eta$ are chosen to be (a) $\eta = 0$, (b) $\eta = 0.2$, (c) $\eta = 0.4$, and (d) $\eta = 0.6$. Dots and error bars correspond to means computed over 1,000 Monte Carlo simulations and the solid lines are the AME predictions.}
    \label{temp_dynamics}
\end{figure*}

To investigate how AME describes the temporal evolution of the system dynamics, especially the transitory dynamics in the early periods, we first present in Fig.~\ref{temp_dynamics} a zoomed version of Fig.~\ref{disease_prevalence} for the region $t \in [0, 1000]$. Overall, AME is able to reproduce the general shape of the time trajectories of the transitory dynamics. In particular, we note AME performs well when initially the network is degree-regular, regardless of $\eta$. For the other two initial degree distributions, AME performs well when $\eta$ is not too large, with the discrepancy between simulations and AME approximations becoming larger as $\eta$ increases, especially for the truncated power law cases.

\begin{figure*}%
    \centering
    \includegraphics[width=2.0\columnwidth]{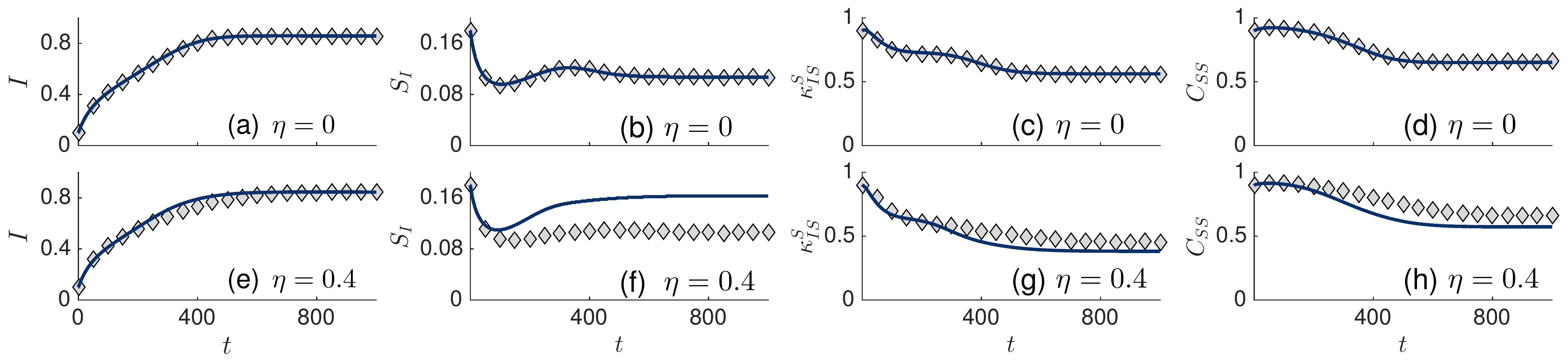}
     \caption{Time evolution of the disease prevalence $I$, the fraction of SI links $S_{I}$, the effective branching factor $\kappa^{S}_{IS}$, and the average number of connections that susceptible nodes share with other susceptible nodes $C_{SS}$. Diamond markers correspond to means computed over 200 Monte Carlo simulations (error bars would be smaller than the markers here) and the solid blue lines are the AME predictions. The initial degree distribution is Poisson and  other parameters of the system are $\beta = 0.04$, $\gamma = 0.04$, $\alpha = 0.005$, $\epsilon = 0.1$, and $\langle k \rangle = 2$. }%
\label{new_measurements}
\end{figure*}

Moreover, inspired by Ref.~\cite{marceau2010adaptive}, in Fig.~\ref{new_measurements}(a-h) we include the evolution of the fraction of $SI$ links $S_I$, the effective branching factor $\kappa^{S}_{IS} \equiv S_{SI}/S_{I}$, and the average number of connections that susceptible nodes share with other susceptibles $C_{SS} \equiv S_{S}/(S_{S} + S_{I})$ as well as the AME approximation for these quantities. In an SIS model,  $S_I$ measures the level of links that could potentially pass the disease, $\kappa^{S}_{IS}$ measures the average number of susceptible neighbors that the infected end of a $SI$ link has, and $C_{SS}$ measures the fraction of susceptible neighbors of a susceptible node. In Fig.~\ref{new_measurements} we observe that AME captures the qualitative behavior of the temporal evolution of the different quantities, but again, the discrepancy between simulations and AME approximations becomes larger as $\eta$ increases.

With it having been shown at multiple places in the main text that AME accurately predicts the stationary level of disease prevalence and various quantities of interest, we also note here the level of agreement between AME and simulation for the stationary levels of the quantities in Fig.~\ref{new_measurements}, with good agreement at $\eta=0$ and increased error for $\eta=0.4$. On the other hand, the differences between simulations at different $\eta$ are not particularly large for $\eta$ up to around $0.6$, and with qualitative differences only for $\eta$=1 (see Figs.~\ref{ER_dynamics} and \ref{disease_time_final}). The changes in the stationary levels predicted by AME do not vary greatly with $\eta$ either, but they appear to do so in the correct direction for these quantities here. While AME does not capture the system dynamics well when $\eta$ is large, it nevertheless provides good predictions for the stationary disease prevalence (Figs.~\ref{disease_time_final}, \ref{bifurcation_plot}, and \ref{bifurcation_2}) and degree distributions (Figs.~\ref{degree_distribution_1} and \ref{degree_distribution_2}) for $\eta$ not too close to $1$.

\end{document}